\documentclass[useAMS,usenatbib]{mnras}
\usepackage{amssymb}
\usepackage{amsmath}
\usepackage{lipsum}
\usepackage{natbib}
\usepackage{multirow}
\usepackage{hhline}
\usepackage[pdftex]{graphicx}
\graphicspath{/Users/mattorr/Desktop/Hopkins/analysis/sfr/losSFR_scripts/nonEQfigs}
\newcommand{\fig}[1]{#1}

\newcommand{\be}{\begin{equation}}
\newcommand{\ee}{\end{equation}}

\newcommand{\Mach}{\mathcal{M}}

\newcommand{\mpcpc}{M$_\odot$ pc$^{-2}$}

\newcommand{\rev}[1]{{#1}}
\newcommand{\tdo}{5}

\newcommand{\dtdo}{30}

\title[Delayed Feedback and SFR Scatter]{A Simple Non-equilibrium Feedback Model for Galaxy-Scale Star Formation: Delayed Feedback and SFR Scatter} %}
\author[M. E. Orr et al.]{Matthew E. Orr$^{1}$\thanks{E-mail: meorr@caltech.edu}, Christopher C. Hayward$^{2}$, 
Philip F. Hopkins$^{1}$ \\
$^{1}$TAPIR, Mailcode 350-17, California Institute of Technology, Pasadena, CA 91125, USA\\
$^{2}$Center for Computational Astrophysics, Flatiron Institute, 162 Fifth Avenue, New York, NY 10010, USA\\}

\begin{document}
\date{Draft date: \today}
\pagerange{\pageref{firstpage}--\pageref{lastpage}} \pubyear{2018}
\maketitle
\label{firstpage}
\begin{abstract}
We explore a class of simple non-equilibrium star formation models within the framework of a feedback-regulated model of the ISM, applicable to kiloparsec-scale resolved star formation relations (e.g. Kennicutt-Schmidt).  Combining a Toomre-Q-dependent local star formation efficiency per free-fall time with a model for delayed feedback, we are able to match the normalization and scatter of resolved star formation scaling relations.  In particular, this simple model suggests that large ($\sim$dex) variations in star formation rates (SFRs) on kiloparsec scales may be due to the fact that supernova feedback is not instantaneous following star formation.  The scatter in SFRs at constant gas surface density in a galaxy then depends on the properties of feedback and when we observe its star-forming regions at various points throughout their collapse/star formation ``cycles".  This has the following important observational consequences: (1) the scatter and normalization of the Kennicutt-Schmidt relation are relatively insensitive to the local (small-scale) star formation efficiency, (2) but gas depletion times and velocity dispersions are; (3) the scatter in and normalization of the Kennicutt-Schmidt relation is a sensitive probe of the feedback timescale and strength; (4) even in a model where $\tilde Q_{\rm gas}$ deterministically dictates star formation locally, time evolution, variation in local conditions (e.g., gas fractions and dynamical times), and variations between galaxies can destroy much of the observable correlation between SFR and $\tilde Q_{\rm gas}$ in resolved galaxy surveys.  Additionally, this model exhibits large scatter in SFRs at low gas surface densities, in agreement with observations of flat outer HI disk velocity dispersion profiles.  %; (5) likewise, the local gas velocity dispersion is well-correlated with the strength of feedback, even if we assume that stellar feedback drives the dispersion. \cch{This last bit sounds awkward to me because of the `even if' -- don't we naively expect such a correlation if feedback 'drives the dispersion'?}
\end{abstract}

\begin{keywords}
galaxies: ISM, evolution, star formation, ISM: kinematics and dynamics.
\end{keywords}
%------------------------------------------------------------------------------------
\section{Introduction}

%Almost anywhere we look in the universe, galaxies are turning gas into stars.  However, these galaxies are not very proficient: the efficiency of the conversion to stars is only a few percent of the gas reservoirs per dynamical time \citep{Kennicutt2007}.
One of the fundamental characteristics of star formation is that it is globally inefficient: galaxies convert only a few per cent of their cold gas reservoirs into stars per dynamical time \citep{Kennicutt2007}.  As to why this is the case, there are two broad frameworks for regulating star formation in galaxies: dynamics and feedback.  Dynamical regulation argues that stars form as rapidly as they are able, but that dynamical processes such as turbulent shear, differential rotation, or gas expansion behind spiral arms govern the fraction of gas with conditions favorable to star formation \citep{Saitoh2008, Robertson2012, Elmegreen2015, Semenov2017}.  In this regime, star formation efficiency (SFE) is low locally, in complement with its global value.  Feedback regulation argues instead that star formation could be locally highly efficient in regions which are actually collapsing without local feedback present, but that stellar feedback (usually in addition to dynamical processes), in the form of ionizing radiation or supernova explosions, heat and stir the interstellar medium (ISM), preventing further star formation in most regions and times \citep[][among others]{Thompson2005, Murray2010, Ostriker2010, Shetty2012, Hopkins2014, Kim2015, Hopkins2017}. 

Within the framework of feedback-regulation there have been several related models describing various star formation `laws', including the ``outer disk" model of \citet{Ostriker2011}, the ``two-zone" theory of \citet{Faucher-Giguere2013}, and radiation pressure supported models like \citet{Thompson2005}, to name a few.  Particular focus has been laid on models involving turbulent support of the ISM, as thermal heating processes become relatively ineffective at regulating star formation for gas surface densities above $\sim$10 M$_\odot$ pc$^{-2}$, where a self-shielded component of the ISM necessarily develops \citep{Schaye2004, Krumholz2009a,Krumholz2009,Hayward2017}.  Broadly, turbulently-regulated models incorporate some metallicity dependence (often having to do with the metallicity dependence of the efficiency of SNe momentum coupling, \citealt{Martizzi2015}), local gas fraction (or stellar surface density, \citealt{Ostriker2011}), or local gas scale height dependence \citep{Faucher-Giguere2013}, in setting the equilibrium star formation rate.

These models have found general agreement with the \emph{mean} observed star formation rates (either galaxy-integrated or as a function of radius) in nearby galaxies.  However, observational studies of the spatially-resolved (at $\sim$kpc scales) Kennicutt-Schmidt relation have apparently-characteristic $\pm  2\sigma$ scatters of $\sim 1-2$ dex in star formation rates at constant gas surface densities \citep{Bigiel2008, Leroy2008, Bigiel2010, Leroy2013, Leroy2017}, with a similar scatter having been seen in cosmological simulations \citep{Orr2018}.  Generally, these variations in star formation rates (SFRs) within individual galaxies at constant gas surface density are not readily explained by local variations in metallicity.  For instance, at fixed galactocentric radii in discs, gas metallicity is seen to vary at $\lesssim 0.1$~dex levels \citep{Ho2017}, whereas gas surface densities can vary by more than 2 dex, requiring SFE~$\propto Z^{20}$ (not seen observationally, or having a theoretical basis for being the case) to explain SFR variations independent of gas surface densities.  Nor are metallicity gradients large enough to explain the scatter, as generally gas surface densities fall far more quickly than metallicities \citep{Ma2017}.   Gas fractions, too, appear lacking in their ability to drive large scatter in SFRs at constant gas surface density \emph{within} galaxies \citep{Leroy2013}.

This large scatter could suggest that we are still missing some critical physics in our models, or observationally our inferred star formation rates and gas surface densities are introducing much larger errors than usually appreciated.  From the side of theory, that we are roughly matching star formation rate distributions, and their scatter in particular, in cosmological simulations is heartening \citep{Orr2018} and suggests the feedback physics included in simulations like those of \citet{Hopkins2014, Hopkins2017} or \citet{Agertz2015} are close to sufficient.  On the side of observations, there remains work to be done in converging on conversion factors between luminosities or line widths, and star formation rates and gas masses but it is unlikely that these factors randomly vary by $\sim2$~dex in neighboring kpc-patches of ISM \citep{Kennicutt2012, Narayanan2012, Bolatto2013}.

Another possible resolution is that rather than star formation being locked to a `law' dependent on gas surface density, there is some ``intrinsic" uncertainty to it \citep{Schruba2010, Calzetti2012, Kruijssen2014, Kruijssen2018}.  \citet{Kruijssen2014} argue that star formation relations like that of the Kennicutt-Schmidt relation must necessarily break down on some scale due to the overlap (or lack thereof) both temporally and spatially between tracers of dense gas and star formation, and that scatter in these relations is a necessary consequence.  But to what extent does the framework of feedback-regulation itself provide an intrinsic scatter to the predicted equilibrium star formation rates?  After all, feedback is not instantaneous with star formation, as ionizing radiation is injected  for upwards of 10 Myr \citep{Leitherer1999}, supernova feedback is not felt for the first $\sim 5$ Myr, and then continues stochastically for $\sim 30$ Myr \citep{Agertz2013}.  The timescales for feedback injection are not coincidentally on the order of the lifetimes of star forming regions themselves in the feedback regulated model \citep{Oklopcic2017, Semenov2018, Grudic2018}.  Star formation \emph{equilibrium} need not be expected, even at the $10^6$ M$_\odot$ giant molecular cloud (GMC) scale. 

%Non-equilibrium star formation on the scales of galaxies \meo{idk segue} Recent work by \citet{Torrey2016} to understand how the outflows and ``breathing modes" of central molecular regions are tied to star formation show that non-equilibrium star formation rates, on timescales of tens of millions of years, arise naturally from the competition between the stellar feedback and dynamical timescales involved. \meo{talk more about galactic centers and their connection to non-equilibrium SFRs..} \meo{yeah a bit of digression here that we have started to think about this in certain contexts, usually quite high star formation rates and gas surface densities..}

Indeed others \citep{Benincasa2016,Torrey2016,Semenov2018} have argued that while star formation might be in ``static equilibrium" (i.e. steady state) in some averaged sense, that it is locally in some \emph{dynamical} equilibrium where the ISM is in a constant cycle of collapse, star formation, and cloud destruction/feedback.  It is thus never instantaneously in local equilibrium, and is constantly oscillating between those phases \citep{Benincasa2016,Semenov2017,Semenov2018}.

In this paper, within the framework of feedback-regulation, we explore a simple non-equilibrium star-formation model, which expands upon these previous works. Critically, we explore models wherein there is a non-trivial delay time, with respect to the local dynamical time, between the formation of young stars and the injection of the bulk of their feedback into the ISM.  We investigate the results of including a time dependence between the criteria for star formation being met, and its effects being felt- in particular, the ability to explain significant ($\sim$dex) scatter in star formation rates in resolved galaxy scaling relations.  We explore how this ultimately leads to scatter in the Kennicutt-Schmidt relation, but also a number of non-intuitive effects for observed galaxy scalings of quantities that enter the model.

\begin{table}\caption{Summary of variables used in this paper}\label{table:variables}
\begin{tabular}{ll}
\hline
 Symbol & Definition  \\
 \hline
$\dot\Sigma_\star$ 		& Star formation rate surface density  \\
$\Sigma_g$ 			& Total gas surface density \\
$f_{\rm sf}$			& Gas mass fraction in star-forming phase \\
$f_g$ 				& Fraction of disk mass in gas \\
$\rho_0$				& Disk mid-plane volume mass density \\ 
$t_d$ 				& Delay timescale for the injection of feedback   \\
$\delta t_d$ 			& Period of feedback injection  \\
$\alpha$ 				& Slope of power law for delay-time distribution of \\
   					&	Type-II SNe   \\
H   					& Gas scale height \\
G   					& Newtonian gravitational constant  \\
$P/m_\star$   			& Characteristic feedback momentum per mass of   \\
   					&	stars formed   \\
$t_{\rm eddy}$			& Eddy (disk scale height) crossing time \\
$\left< \epsilon_{\rm sf} \right>$ 		& Average star formation efficiency per eddy time\\
					& (here, GMC-scale average value) \\
$\bar \epsilon_{\rm sf}$ 	& Star formation efficiency per orbital \\
					& dynamical time\\
$\tilde Q _{\rm gas}$			& Modified Toomre-Q gas stability parameter \\
$\Omega$				& Local orbital dynamical time \\
$\sigma$				& Turbulent gas velocity dispersion (3-D) \\

  %. ....plenty more variables to define
 \hline
\end{tabular}
\end{table}
%------------------------------------------------------------------------------------
\section[]{Model}\label{mod}

In a previous work \citep{Orr2018}, we explored the ability of turbulent energy injection, in the form of the effects of Type II SNe, to explain the equilibrium value of the Kennicutt-Schmidt relation in the FIRE simulations at gas surface densities $\gtrsim 10$ M$_\odot$ pc$^{-2}$ \citep[similar in derivation to][]{Ostriker2011,Faucher-Giguere2013, Hayward2017}.  The predicted equilibrium was in good agreement with the median values seen in the simulations, which were themselves in good agreement with the observed atomic+molecular formulation of the Kennicutt-Schmidt relation.  However, the $\pm 2\sigma$ scatter seen, on the order of $\sim 1.5-2$~dex, was not fully explained by local environmental variations, e.g. metallicity, dynamical time, or stellar surface density.  There appeared to be an intrinsic scatter of $\gtrsim$dex to the star formation rate distribution seen at any given gas surface density.

To explore the physical processes that cause scatter in resolved star formation scaling relations in disk environments within individual galaxies, let us consider a patch of the ISM where the turbulent velocity dispersion is taken to be roughly isotropic, where we assume
\begin{equation} 
\sigma^2 = \sigma_R^2 + \sigma_z^2 + \sigma_\phi^2 \approx 3 \sigma_R^2 \; ,
\end{equation}
or $\sigma \approx\sqrt{3}\sigma_R$ where $\sigma$ is the overall gas velocity dispersion, and the subscripted $\sigma$'s denote the velocity dispersions in the radial, vertical (i.e. line of sight in face-on galaxies), and tangential directions, respectively.

In the framework of a supersonic turbulent cascade, the largest eddies carry the bulk of the energy and momentum to first order, and we can take the momentum per area in the turbulent/random motion of the gas to be the velocity dispersion at the largest scale (here, the gas disk scale height $H$) times the gas mass surface density $\Sigma_g$, that is $P_{turb} = \Sigma_g \sigma$.  The timescale for the dissipation $t_{diss}$ of this turbulent momentum\footnote{In \citet{Faucher-Giguere2013}, they assume that turbulent \emph{energy} dissipates in an eddy crossing time.  However, if $E_{turb} \sim P_{turb}^2/2\Sigma_g$ and $\Sigma_g$ is constant, then $\dot E_{turb} \sim P_{turb}\dot P_{turb}/\Sigma_g$. The exponential turbulent energy dissipation rate $\dot E_{turb} \sim -E_{turb}/t_{\rm eddy}$ becomes $P_{turb}\dot P_{turb}/\Sigma_g \sim -P_{turb}^2/2\Sigma_g t_{\rm eddy}$, reducing to $\dot P_{turb} \sim -P_{turb}/2t_{\rm eddy}$, i.e., that the turbulent \emph{momentum} decays approximately in twice an eddy crossing time.  For consistency, and since SNe are momentum-conserving, we adopt a momentum-centric focus throughout the paper.} is roughly the (twice) eddy turnover time $t_{\rm eddy}$, which is $t_{\rm eddy} \approx H/\sigma_z$.  If we assume that the gas disk is embedded in the potential of stellar disk with a larger scale height, as is seen in the Milky-Way with the thin gas disk having a characteristic height of $\sim100$~pc embedded within the larger $\sim300$~pc stellar scale height \citep{Gilmore1983, Scoville1987}, and that the gravitational acceleration near the mid-plane due to the local disk mass itself is of the form $4 \pi G \rho_0 z$, where $\rho_0$ is the mid-plane density (gas + stars), and the external potential\footnote{Here, the local dark matter contribution is implicitly included, whereas it is ignored for simplicity in the disk self-gravity acceleration term as the baryonic component dominates the thin disk mass in galaxies.  Our model could be extended to gas-rich dwarfs or high-redshift galaxies with poorly defined disks, but would require a different formulation of gas scale-lengths/heights.} introduces a vertical acceleration component of $v_c^2z/R^2 = \Omega^2 z$ (where $\Omega \equiv v_c/R$), then the vertical (z) density profile is a Gaussian with a characteristic scale height of
\begin{equation}
H = \frac{\sigma_z}{\Omega + \sqrt{4 \pi G \rho_0}} \; .
\end{equation}
So, $t_{diss} \approx 2t_{\rm eddy} \approx 2H/\sigma_z \approx 2/(\Omega + \sqrt{4 \pi G \rho_0})$.  In the absence of stellar feedback, the turbulent momentum of this patch of the ISM would be expected to exponentially decay as
\begin{equation} \label{eq:turb_decay}
\dot P_{turb} = -\Sigma_g \sigma/t_{diss} = -P_{turb}(\Omega+\sqrt{4 \pi G \rho_0})/2\; ,
\end{equation}
which admits a solution for gas velocity dispersions of $\sigma(t) = \sigma_0 \exp{(- t(\Omega + \sqrt{4 \pi G \rho_0})/2)}$.

%\subsubsection{Connection to Isolated Disk Models with No-Feedback and the Kennicutt-Schmidt Relation}
%A number of studies have explored how the Kennicutt-Schmidt relation, nominally taken to be a description of how gravity and feedback compete, arises in simulations and models without any explicit baryonic feedback. The star formation physics explored by those works is key to why a number of them find a correctly-normalized no-feedback Kennicutt-Schmidt relation.
%
%We consider models of star formation in which stars form from dense gas with a given efficiency $\epsilon_0$ per free fall time $t_{ff}$ (nearly any timescale will do, so long as it is shorter than the dynamical time of the system, $1/\Omega$), above a set density threshold $n_{\rm crit}$, i.e. the star formation rate is,
%\begin{equation}
%\dot M_\star = \epsilon_0 M_g(n > n_{\rm crit})/t_{ff} \: ,
%\end{equation}
%where $M_g(n > n_{\rm crit})$ is the mass of gas above $n_{\rm crit}$.  If $n_{\rm crit}$ is sufficiently high, with `very dense' star forming gas making up only a small fraction of the total gas budget, star formation is rate limited not by feedback
%
%
% \meo{write up a short model incorporating a log-normal sub-patch density pdf, which forms stars above a threshold density. Follow that SFR and show we get a KS relation.}

\subsection{Equilibrium Model of Instantaneous Feedback Injection in Disk Environments}\label{ss:eq_sfr}
However, feedback from massive stars acts to inject momentum back into the ISM at the largest scales \citep[i.e. disk scale heights,][]{Padoan2016}.  %Following \citet{Ostriker2011, Faucher-Giguere2013},
Taking the characteristic momentum injected per mass of young stars formed to be $P/m_\star$, we can establish an equilibrium for $\sigma$ if we balance the rate of momentum injection from feedback, $\dot\Sigma_\star P/m_\star$, with the turbulence dissipation rate in Eq.~\ref{eq:turb_decay}, that is,  
\begin{equation}\label{eq:sfr_balance}
\left(\frac{P}{m_\star}\right)\dot \Sigma_\star = \Sigma_g \sigma  (\Omega+ \sqrt{4 \pi G \rho_0})/2 \; .
\end{equation} 
Arguing that star-forming disks are marginally stable against gravitational instabilities, we invoke a modified\footnote{This is not the `real' two component Toomre-Q \citep{Rafikov2001}, but is a much simplified version that is sufficiently accurate for our purposes (using the full two-component Q makes little difference to our numerical calculations but prevents us from writing simple analytic expressions).} Toomre-Q criterion dictating instantaneous gas stability \citep{Toomre1964},
\begin{equation}\label{eq:Q}
\tilde Q_{\rm gas} = \frac{\sqrt{2}\sigma_R \Omega}{\pi G \Sigma_{disk}} \; ,
\end{equation}
where $\Sigma_{\rm disk} = \Sigma_g + \gamma\Sigma_\star$ is the mid-plane surface density, including the stellar component (with the factor $\gamma$ accounting for the effective fraction of stellar mass within a gas scale height, $\gamma = 1 - \exp(-H/H_\star)$).  We substitute this Toomre-Q into Eq.~\ref{eq:sfr_balance} for $\sigma$, recovering the Kennicutt-Schmidt relation for a turbulently supported ISM,
\begin{equation} 
\dot \Sigma_\star = \pi G \tilde Q_{\rm gas} \sqrt{\frac{3}{8}}  \frac{\Sigma_g \Sigma_{disk}}{P/m_\star}\left( 1 + \frac{\sqrt{4 \pi G \rho_0}}{\Omega}\right) \; .
\end{equation}
Further, we can calculate the ``global star formation efficiency", i.e. the fraction of the gas mass converted to stars per orbital dynamical time, $\bar\epsilon_{\rm sf} \equiv \dot\Sigma_\star/\Sigma_g\Omega$, to be
\begin{equation} 
\bar\epsilon_{\rm sf} = \pi G \tilde Q_{\rm gas} \sqrt{\frac{3}{8}}\frac{\Sigma_{disk}(\Omega+ \sqrt{4 \pi G \rho_0})}{(P/m_\star) \Omega^2} \; .
\end{equation}

If we take $\tilde Q_{\rm gas}$ to be a constant, assuming a value near or slightly below one, and consider the case in which the disk is not strongly self-gravitating (likely, with the marginal stability of $\tilde Q_{\rm gas}\approx1$), such that $\Omega >> \sqrt{4 \pi G \rho_0}$; these two relations boil down to a description of gas surface density and mass fraction and a representation of the ratio of disk surface density to inverse dynamical time, respectively:
\begin{equation} \label{eq:equilibs}
\dot \Sigma_\star = \pi G \sqrt{\frac{3}{8}}  \frac{\Sigma_g \Sigma_{disk}}{P/m_\star} \; \; \& \; \; \bar\epsilon_{\rm sf} =\pi G \sqrt{\frac{3}{8}}  \frac{ \Sigma_{disk}}{\Omega P/m_\star}.
\end{equation}

One deficiency of this model of feedback regulation lies in the calibration of the strength of feedback to isolated Type-II SNe simulations \citep[e.g.,][]{Kim2015b, Martizzi2015}.  Generally, this overlooks the variation in effective feedback coupling due to the local environment.  Especially for predictions regarding the line of sight velocity dispersions, the potential saturation or ``venting" of feedback after SNe remnants (super-bubbles or otherwise) break out of the disk plane \citep{Fielding2017}, or the enhanced momentum injection efficiency of spatially-clustered SNe \citep{Gentry2019}, are possible concerns.  We do not explore the effects of feedback saturation or SNe (spatial) clustering here, but they warrant further exploration within the framework of simple analytic models (these effects are self-consistently handled in galaxy simulations that resolve gas disks and supernova remnants in the snowplow phase).

\subsection{Non-equilibrium Model of Feedback Injection in Disk Environments}\label{ss:noneq_sfr}
The model derived in \S\ref{ss:eq_sfr} is an equilibrium model, which assumes that feedback injection is statically balanced with the dynamical/dissipation rate.  However, we might consider here that the departures from equilibrium occurring on the feedback delay timescale are important for setting the scatter seen in $\dot\Sigma_\star$ at constant $\Sigma_g$ in the Kennicutt-Schmidt relation, and at constant $\Sigma_g \Omega$ for the Elmegreen-Silk relation, as well as in $\sigma_z$--$\dot\Sigma_\star$ space.  We will explicitly consider only delayed feedback (i.e. Type-II SNe) in this model.\footnote{Although prompt feedback (e.g. radiation pressure and stellar winds) injects a similar amount of momentum per mass of young stars over their lifetimes \citep{Agertz2013}, the `characteristic' velocity at which this momentum couples to the ISM on large scales is lower by a factor of 20 or so, compared to SNe feedback \citep{Murray2010, Faucher-Giguere2013}.  As we consider here the ability of feedback to regulate the disk scale properties that regulate star formation `from the top down', we neglect explicitly treating the prompt feedback effects in our model.  Instead, we implicitly incorporate its effects regulating the efficiencies of cloud-scale, $<100$~pc, star formation in our ``GMC-scale" star formation efficiency model \citep{Grudic2018}.} 

Rather than holding the turbulent velocity dispersion $\sigma$ constant in time, we allow it to vary, defining the behavior of its derivative $\dot\sigma$ as,
\be \label{eq:sigma_evo}
\dot\sigma = \dot\sigma_{\rm SNe} - \sigma/t_{\rm eddy} \; ,
\ee
where $\dot\sigma_{\rm SNe}$ is the term explicitly following the current injection of SNe feedback momentum due to past star formation (see Eq.~\ref{eq:sig_sne}, below), and the $\sigma/t_{\rm eddy}$ term accounts for the exponential decay of supersonic turbulence on roughly an eddy crossing time (Eq.~\ref{eq:turb_decay}).  We ignore the fraction of turbulent momentum ``locked away" into stars (equivalent to a $\sigma \dot\Sigma_g$ term) as the term is negligible with the depletion time of gas typically on the order of $\sim$Gyr in galaxies \citep{Leroy2008, Leroy2013}.

Developing a form for $\dot\sigma_{\rm SNe}$, we consider that Type-II SNe feedback from a given star formation event is injected after a delay time $t_d$, and over a period $\delta t_d$, corresponding to the lifetime of the most massive star formed, and the time until the least massive star to undergo core-collapse does so thereafter.  Furthermore, convolving the number of stars of a given mass with their lifetimes produces a shallow power-law distribution in time over which  SNe occur after a star formation event, such that $dN_{SNII}/dt \propto t^{-\alpha}$ (see Appendix~\ref{append:SNe} for a more detailed derivation).  These quantities, $t_d$, $\delta t_d$, and $\alpha$, are reasonably known (see Appendix~\ref{append:SNe}), and we adopt fiducial values in this paper of 5 Myr, 30 Myr, and 0.46, respectively.  As such, the governing equation for $\dot\sigma_{\rm SNe}$ takes the form
\be \label{eq:sig_sne}
\Sigma_{\rm g}\dot\sigma_{\rm SNe} = (P/m_\star) \chi \int^{t_d + \delta t_d}_{t_d} \frac{\dot\Sigma_\star(t-t')}{t'^{\alpha}} dt' \; ,
\ee
where $P/m_\star$ here is the momentum injected by Type-II SNe event per mass of young stars (as opposed to from all sources of feedback as in \S~\ref{ss:eq_sfr}), and $\chi$ is a normalization factor such that for a constant star formation rate $\dot\Sigma_\star$ the equation reduces to $\Sigma_{\rm g}\dot\sigma_{\rm SNe} = (P/m_\star) \dot\Sigma_\star$.  We adopt a fiducial value of $P / m_\star = 3000 $~km/s \citep[the same value adopted by the FIRE simulations of ][]{Hopkins2014, Hopkins2018:fire}, and explore the effects of varying the strength of SNe feedback in \S~\ref{res:fb_str}.

It is then necessary to formulate a model for the rate at which star formation proceeds, as a function of the current state of the ISM, as we now consider $\dot\Sigma_\star$ to drive $\dot\sigma$, rather than being purely in a static equilibrium with the turbulent dissipation.  

Taking the large-scale marginal gas stability as a key parameter in setting the current rate of star formation, we invoke a simple ``two-phase" model of the ISM, which is instantaneously dependent on the Toomre-Q parameter of the gas disk.  Let us assume that some fraction of the gas is in a star-forming phase $f_{\rm sf}$ (i.e. marginally gravitationally-bound gas), with the remaining mass in a non-star-forming phase. As explored analytically by \citet{Hopkins2013a}, supersonic turbulence drives parcels of gas to randomly walk in log-density space such that a fraction (here, $f_{\rm sf}$) are driven to sufficient densities such that local collapse (i.e. leakage) occurs even if the global value of $\tilde Q_{\rm gas}$ exceeds the critical threshold for gravitational instabilities $Q_0$\footnote{This is just a formal calculation of the log-normal density distribution of gas in supersonic turbulence.  It is to say: turbulence is able to dynamically replenish the fraction of gas in a log-normal density distribution that is above some critical threshold for self-gravity and collapse.}.  Following the rationale of \citet[][see their Appendix C]{Faucher-Giguere2013}, adapting the calculations of \citet{Hopkins2013a}, we argue that the mass fraction of gas susceptible to gravitational collapse ($f_{\rm sf}$), which subsequently would be considered in some stage of ``star-forming", is functionally dependent on Toomre-Q, with an adopted power-law form of,
\be\label{eq:logeff}
f_{\rm sf}(\tilde Q_{\rm gas}) = f^0_{\rm sf}\left(\frac{Q_0}{\tilde Q_{\rm gas}}\right)^{\beta}\; , 
\ee
for values $\tilde Q_{\rm gas} > Q_0$, and is a constant $f^0_{\rm sf}$ for $\tilde Q_{\rm gas} < Q_0$, where $f^0_{\rm sf}$ is the maximal fraction of gas in the star-forming phase, $Q_0$ represents the Toomre-Q stability threshold, and $\beta$ accounts for the ``stiffness" of that threshold.   Further, as $\tilde Q_{\rm gas}$ evolves (in this model, through evolution purely in $\sigma$) smoothly in time, the roll-on (or off, if $\dot\sigma > 0$) can also be thought to implicitly parameterize our ignorance in how and at what rate GMCs assemble (for $\dot\sigma > 0$, this can approximate ionizing radiation and winds dispersing dense material).  In \citet{Hopkins2013a}, the stiffness of the instability threshold ($\sim \beta$, here) was inversely dependent on the Mach number  $\Mach$ of the turbulence-- intuitive, as larger Mach numbers yield a broader log-normal density distribution, increasing the amount of gas above a given density relative to the mean gas density, hence softening the effective gravitational instability threshold.  Here, taking $\Mach \sim \sigma/c_s$, where $c_s$ is the speed of sound for $\sim 300$~K molecular gas, and $\tilde Q_{\rm gas} \sim$ constant, we thus have $\Mach \propto \sigma \propto \Sigma_g$.  And so, in our model at a given gas surface density we adopt a stiffness $\beta = -2 \log( \Sigma_g/{\rm M_\odot pc^{-2}}) +6$, proportional to the Mach number-dependent stiffness fit by \citep{Faucher-Giguere2013}, and substantiated by the observational findings relating $\Sigma_g$ and $\Mach$ of \citet{Federrath2017}.

Arguing that a $\sim$kpc-sized patch of the ISM likely incorporates a large enough number of $\lesssim$100~pc clouds so as to approach an average behavior in terms of their individual evolutionary states \citep{Schruba2010, Calzetti2012, Kruijssen2014}, we then adopt a $\sim$kpc-scale star formation rate of
\be \label{eq:sfr_Q}
  \dot\Sigma_\star(t) = \left< \epsilon_{\rm sf} \right> f_{\rm sf}(\tilde Q_{\rm gas}(t))\Sigma_g/t_{\rm eddy}
\ee
where $f_{\rm sf}(\tilde Q_{\rm gas}(t))\Sigma_g$ is the mass of gas in the star-forming state (per area), $\left< \epsilon_{\rm sf} \right>$ is the average star formation efficiency per eddy-crossing time \citep[fiducially, 0.025, in line with cloud-scale efficiencies discussed in ][]{Elmegreen2018}, and $t_{\rm eddy}$ is the eddy-crossing time.  As the quickest instabilities to grow are at the largest scales, the largest being that of the disk scale height itself, the effective free-fall time of gas at the mid-plane density is equivalent to the eddy crossing time $t_{\rm eddy}$ up to an order unity factor (since $t_{ff} \sim 1/\sqrt{G\rho_0} \sim t_{\rm eddy}$).  Again, emphasizing that we defined our efficiency $\left< \epsilon_{\rm sf} \right>$ (taken to be a constant) as a kpc-scale average quantity, $\left< \epsilon_{\rm sf} \right> \equiv \left<\dot M_\star t_{\rm eddy}/M_{\rm GMC} \right>$ where $M_{\rm GMC} = f_{\rm sf}(\tilde Q_{\rm gas}(t))M_g$.  It is analogous to a GMC-scale average star formation efficiency, and as such is unable to distinguish between high or low efficiency star formation modes on smaller scales \citep[e.g., efficiencies calculated on the basis of higher density gas tracers like HCN][]{Kauffmann2017,Onus2018}.

The fiducial values of the physical quantities and common initial conditions included in the evolution of our model-- essentially the behavior of the PDE for $\sigma$, Eq.~\ref{eq:sigma_evo}, are enumerated in Table~\ref{table:fidmodel}.  The initial condition of the gas in the model, in all cases presented here, is taken to be $\tilde Q_{\rm gas}(t=0) = Q_0 +1$ (and its corresponding velocity dispersion $\sigma$) for the given $\Sigma_g$, embedded within static stellar disk with thin and thick components having scale heights of 350 and 1000~pc, respectively, and a relative mass fraction $f_{\rm thick} \equiv \Sigma_{\rm thick,\star}/(\Sigma_{\rm thick,\star}+\Sigma_{\rm thin,\star}) = 0.33$.

\begin{table}\caption{Fiducial Model Parameters and Disk Conditions}\label{table:fidmodel}
\begin{tabular}{lll}
\hline
Parameter & Quantity & Fiducial Value  \\
 \hline 

Toomre-Q Threshold 		& $Q_0$ 					& 1.0  	\\
Max. star-forming fraction 		& $f^0_{\rm sf}$			& 0.3		 \\
Average SF efficiency 		& $\left< \epsilon_{\rm sf} \right>$			& 0.025	  \\
Feedback Strength			& $P/m_\star$				& 3000 km/s    \\
Feedback Delay Time		& $t_d$ 					& 5 Myr  \\
Feedback Duration			& $\delta t_d$ 				& 30 Myr	\\
Power law slope of Type-II 	&$\alpha$ 				& 0.46  \\
SNe delay time distribution 	&						&\\
Orbital Dynamical Time		& $\Omega$				& 35 Gyr$^{-1}$ \\
Disk Gas Fraction			& $f_g $ 					& 0.33 \\
Stellar Thick Disk Fraction 	& $f_{\rm thick}$ 			& 0.33 \\
Stellar Disk height (thin) 		& $H_{\rm thin, \star}$		& 350 pc  \\
Stellar Disk height (thick) 		& $H_{\rm thick, \star}$		& 1000 pc	\\
  %. ....plenty more variables to define
 \hline
\end{tabular}
\end{table}

\subsubsection{Connecting $\dot \Sigma_\star$, $\Sigma_g$ with Observables}\label{sec:mockobs}
Except for the nearest star forming regions, (where young star counts or protostellar cores can be used as proxies), observers rarely have true estimates for the `instantaneous' star formation rate of a star forming region.  As such, we must connect our `instantaneous' star formation rate with observables like H$\alpha$ or IR flux, which are used as average measures of star formation over a recent period of time $\sim 2-4$~Myr.  For this reason, when we make attempts to compare with observational star formation relations, we average the instantaneous star formation rate $\dot \Sigma_\star$ over the last 3 Myr (see Appendix~\ref{res:times} for how our results vary with the averaging window).  %This particular weighting is justified by the relative flux of H$\alpha$ photons from populations of young massive stars at various ages between 0-10 Myr \citep{} \meo{who was that again?  ... aka how the fuck do I weight things for the H$\alpha$ proxy?}. %who was that again?
To compare our gas surface densities with observations, we take our gas mass surface density $\Sigma_g$ to be the atomic+molecular hydrogen gas, correcting them for Helium mass with a factor of 0.75.  %To connect with the observationally inferred surface densities, we take the natural area of our ``pixels" to be that of the gas scale height at any time, i.e. when presenting our results we will take $\dot\Sigma_\star \equiv \dot M_{\star,avg}/(\pi H^2)$ and $\Sigma_{\rm gas} \equiv M_g/(\pi H^2)$. 

In panels where we plot the Kennicutt-Schmidt relation, we compare results of our simple model with resolved Kennicutt-Schmidt observations from \citet{Bigiel2008} (light and dark grey shaded regions in background).  We correct the gas surface densities in their data with a variable $X_{\rm CO}$ fit from \citet{Narayanan2012}.  Where we plot depletion time against gas stability (Toomre-Q), we compare with the results of \citet{Leroy2008} (light and dark grey shaded regions in background).  For the gas velocity dispersion--star formation rate panels, we present data from the SAMI IFU survey of kpc-scale resolved observations of star forming disks of \citet{Zhou2017}.  As well, we include HI velocity dispersion data of spiral disks from \citet{Ianjamasimanana2015} from the THINGS survey.  These data correspond to velocity dispersion--gas surface density observations, lacking direct SFR data.  However, given that they are at low gas surface density ($\Sigma_g < 10$ M$_\odot$ pc$^{-2}$), we take their results to correspond to a range of SFRs for the low gas surface density region in the \citet{Bigiel2008} dataset.  They are thus presented as a $5-12$ km/s band ranging in $\log(\dot\Sigma_\star/{\rm M_\odot \, yr^{-1} \, kpc^{-2}})$ from -2 to -5, constraining the low velocity dispersion, low-SFR region for our models.
 
%------------------------------------------------------------------------------------
\section{Results}\label{results}
\begin{figure}
	\centering
	\includegraphics[width=0.42\textwidth]{\fig{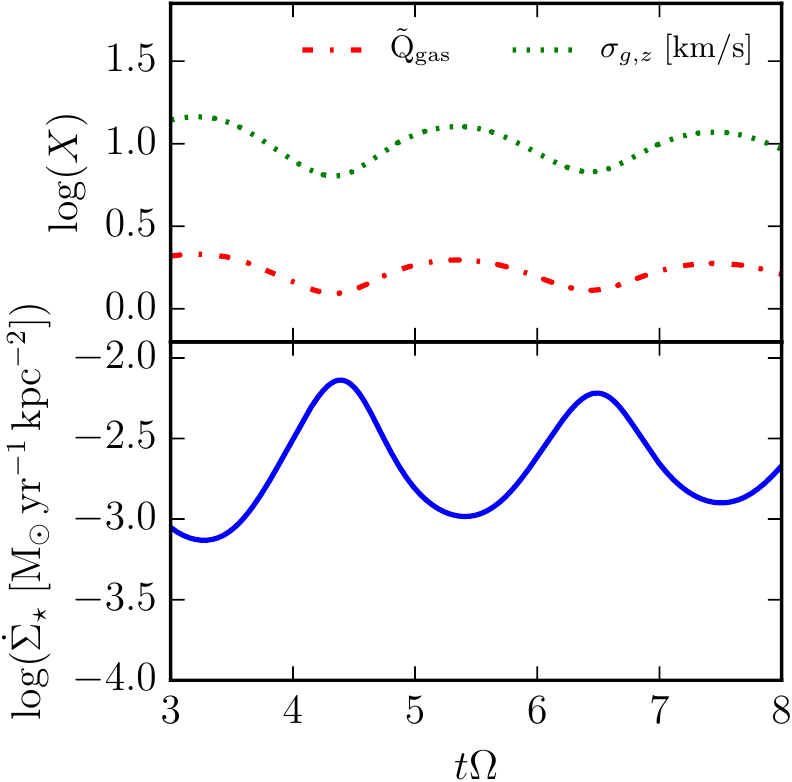}}
	\caption{Logarithmic values of star formation rate surface density (solid blue line; 3-Myr-averaged rate), local $\tilde Q_{\rm gas}$ (dash-dotted red line), and gas velocity dispersion (dotted green line, units: km/s) for a period of five dynamical times in our fiducial model gas patch (for fiducial model parameters, see Table~\ref{table:fidmodel}) with $\Sigma_g=15$ M$_\odot$ pc$^{-2}$ and $\Sigma_\star =35$ M$_\odot$ pc$^{-2}$. The SFR and velocity dispersion maintain stable, albeit slowly decaying, cycles after approximately one dynamical time $\tau_{\rm dyn} \sim \Omega^{-1} \sim 30$~Myr.} \label{fig:fidpatch}
\end{figure}

\begin{figure*}
	\centering
	\includegraphics[width=0.95\textwidth]{\fig{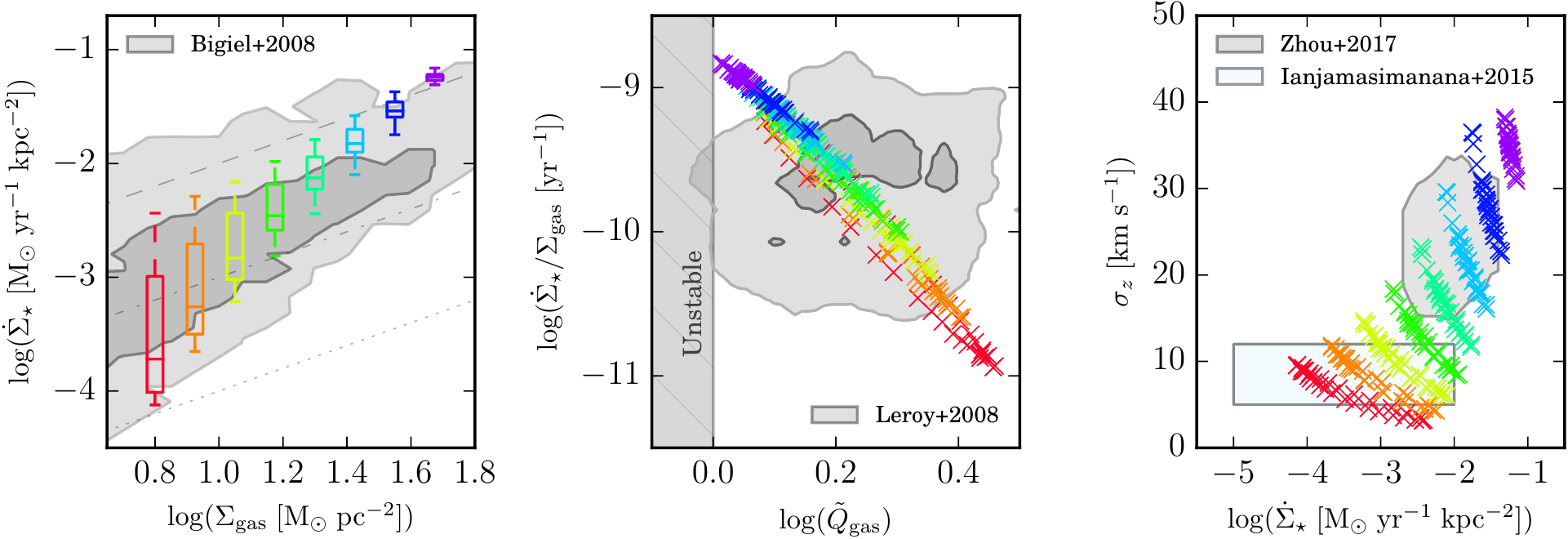}}
	\caption{Fiducial model Kennicutt-Schmidt {\bf(left)}, gas depletion time---Toomre-Q  {\bf(middle)}, and gas velocity dispersion---SFR {\bf(right)} relations for the fiducial parameters listed in Table~\ref{table:fidmodel}. The shaded regions in the background represent observational data ranges (c.f. \S~\ref{sec:mockobs}) from \citealt{Bigiel2008} (left panel), \citealt{Leroy2008} (center panel), and \citealt{Ianjamasimanana2015} and \citealt{Zhou2017} (light blue and grey, respectively, right panel). The dashed, dot-dashed, and dotted lines in the KS panel indicate constant depletion times of 10$^9$, 10$^{10}$, and 10$^{11}$ yr, respectively.  The hatched grey shaded region to the left in the middle panel denotes the Toomre-unstable region.  The fiducial model exhibits good agreement with observations of Kennicutt-Schmidt and gas velocity dispersions.  The Q-threshold is sufficiently soft with its $f_{\rm sf}(\tilde Q_{\rm gas})$ `leakage' to allow star formation to reverse collapse before reaching $Q_0$/disk instability itself.  The upturn in $\sigma_z$--SFR above $\dot\Sigma_\star \approx$~10$^{-2}$ M$_\odot$ yr$^{-1}$ kpc$^{-2}$ reflects the fact that feedback from individual star formation events injects a smaller fraction of the overall ISM turbulent momentum and thus is less effective at changing the gravitationally-unstable fraction of the ISM (especially true, given that the model lacks outflows to remove gas).} \label{fig:fidKS}
\end{figure*}
%\begin{figure}
%	\centering
%	\includegraphics[width=0.45\textwidth]{\fig{nonEQmodel_varFB_rad+winds.pdf}}
%	\caption{Effects on star formation rates and gas velocity dispersions of variation in the strength and duration of the prompt feedback component, in two model patches with the same conditions as those in Figure~\ref{fig:varFB_SNe}.  Plotted quantities and observational data regions are in the style of Figure~\ref{fig:varFB_SNe} as well. {\bf (Left column)} The fraction $\eta$ of the overall feedback ($P/m_\star$) delivered as ``radiation pressure and winds" over the period $t_w$ (=7.5 Myr, fiducial value) sets the high end of the velocity dispersion range, and where the SFRs begin to fall to their minimal values in the KS relation. {\bf (Right column)} The timescale over which the radiation pressure and winds are injected $t_w$, sensitively affects the ability of the model to depart from SF equilibrium.  The shorter periods for feedback injection force the star-formation response closer to equilibrium.}\label{fig:varFB_rad+winds}
%\end{figure}

The simple model produces relatively stable cycles of star formation, inflation and decay of gas velocity dispersions, and variation in the values of the Toomre-Q parameter, as seen in Figure~\ref{fig:fidpatch} for our set of fiducial values of physical parameters, with disk surface densities and conditions chosen to match the solar circle \citep[$\Sigma_g =15$ M$_\odot$ pc$^{-2}$, $\Sigma_\star= 35$ M$_\odot$ pc$^{-2}$, and $\Omega =35$ Gyr$^{-1}$][]{McKee2015}.  As star formation is slow and inefficient (gas depletion times are $\gtrsim$~Gyr here), and given the fact that we do not include some gas outflow term, we do not allow $\Sigma_g$ or $\Sigma_\star$ to vary in the model.  And so, $\tilde Q_{\rm gas}$ and $\sigma_z$ are in phase throughout their cycles, by definition since $\tilde Q_{\rm gas} \propto \sigma_z$ here, ignoring the relatively weak sigma-dependent $\gamma$ term in front of $\Sigma_\star$ in $\Sigma_{\rm disk}$.  Moreover, given the relative stiffness of the star formation threshold in Toomre-Q (for $\Sigma_g = 15$ M$_\odot$ pc$^{-2}$, the `stiffness' of $f_{\rm sf}(\tilde Q_{\rm gas})$ is $\beta \sim 4.6$), star formation commences and is arrested by feedback before $\tilde Q_{\rm gas}$ reaches $Q_0 (=1)$, after which the delayed effects of feedback play out, driving $\tilde Q_{\rm gas}$ and the velocity dispersions to their maximal values before the cycle starts anew.  The instantaneous star formation rate (not shown) is nearly completely out of phase with the velocity dispersions and Toomre-Q, rising sharply as $\tilde Q_{\rm gas}$ falls and falls nearly as quickly as it rises.  The ``observable" quantity, the 3~Myr averaged star formation rate (c.f. the H$\alpha$ SFR tracer), shows how the ``observed" star formation rates rise by $\sim$dex as $\tilde Q_{\rm gas}$ approaches its minimal value, before falling as the effects of SNe feedback are felt later in the star formation episode.

Variations in the overall strength of feedback, the timing of feedback, and star formation prescription all affect the shape and magnitudes of the star formation cycles in the model, but largely the aforementioned picture holds so long as the timescale of feedback relative to the dynamical time of the system is short but not effectively instantaneous, and that the magnitude of feedback is insufficient to totally disrupt the system.  This therefore applies to both galactic centers and in the outskirts of disks, even where the dynamical time is quite long compared to feedback timescales, so long as the ISM is turbulently regulated.

Figure \ref{fig:fidKS} shows the extent of the star formation cycles in the fiducial model across $\sim$dex in $\Sigma_g$ in the Kennicutt-Schmidt, depletion time---stability, and star formation rate---gas velocity dispersion relations.  Results in this figure, and throughout the paper, are plotted as box-and-whiskers in the KS panel represent the median, interquartile region, and 5-95\% data range of individual models run at a given $\Sigma_g$.  \rev{Figure~\ref{fig:fidKS} was run for a range in $\log\Sigma_g = 0.8 -1.675$ with $\log\Sigma_g$ steps of 0.125 dex, all other figures use a range of $\log\Sigma_g = 0.8 -1.55$ with 0.25 dex $\log\Sigma_g$ steps, where $\Sigma_g$ is expressed in units of M$_\odot$ pc$^{-2}$.} Points in other panels (gas velocity dispersion and depletion time--stability relations) are sampled time-steps from those models (seen as clearly separated families of colored points in right panel of Figure~\ref{fig:fidKS}).

At low $\Sigma_g$, the model exhibits increasingly large scatter\footnote{Regions in an ``off"/low-SFR mode of the cycle may likely be counted as entirely non-star forming in observations, dependent on flux thresholds, given their very low SFRs.} as the effects of feedback from peak star formation rates contribute significantly to the overall momentum budget of the disk (c.f. \S~\ref{res:low}), producing a larger scatter to in SFRs for KS, and a spur to long depletion times and `high' Toomre-Qs.  In $\sigma$--$\dot\Sigma_\star$ space, this is seen as a flattening of the relation, covering broad ranges in $\dot\Sigma_\star$ with little change in $\sigma$.  This is broadly in agreement with observations of HI disks in galaxy outskirts having flat velocity dispersion profiles \citep{Ianjamasimanana2012, Ianjamasimanana2015}. The large velocity dispersions in gas seen above $\dot\Sigma_\star \approx$~10$^{-2}$ M$_\odot$ yr$^{-1}$ kpc$^{-2}$ reflect the fact that feedback is simultaneously able to drive outflows and turbulence in the cold ISM at these SFRs \citep{Hayward2017}.  However, in a multiphase ISM, these high dispersions $\sigma_z$ would not appear in the cold ISM turbulence as this feedback would instead drive outflows (and thus dispersions in the warm neutral and ionized gas components).

Counter-intuitively -- but of central importance to observers -- when this model is applied to galaxies as a whole (i.e. many $\lesssim$~kpc patches), the relatively tight correlation between Toomre-Q (or gas $\sigma_z$) and resolved star formation rates within individually evolving patches may be smoothed out by variations in e.g., local gas fractions, dynamical times, star formation efficiencies, or strength of feedback (i.e., the amount of momentum \emph{coupled} into the cold phase of the ISM per mass of young stars), which may shift subsets of the distribution (c.f., later sections of this paper), effectively widening it on galaxy scales to the relatively broad distribution observed by \citet{Leroy2008}.  This argument holding for $\dot\Sigma_\star \lesssim$~10$^{-2}$ M$_\odot$ yr$^{-1}$ kpc$^{-2}$, above which outflows would be possible, the presence of which may affect interpretations of distributions in depletion time--Toomre-Q \citep[and $\sigma_z$ here would no longer strictly encapsulate turbulence in the cold ISM,][]{Hayward2017}. %This is largely due to the fact that variations in Q (or $\sigma_z$) are on the order of a factor of two or so, whereas star formation rates will typically vary by more than an order of magnitude, wiping out their direct connection.

\subsection{Variations in the Strength and Timing of Feedback}\label{res:fb_str}

\begin{figure*}
	\centering
	\includegraphics[width=0.95\textwidth]{\fig{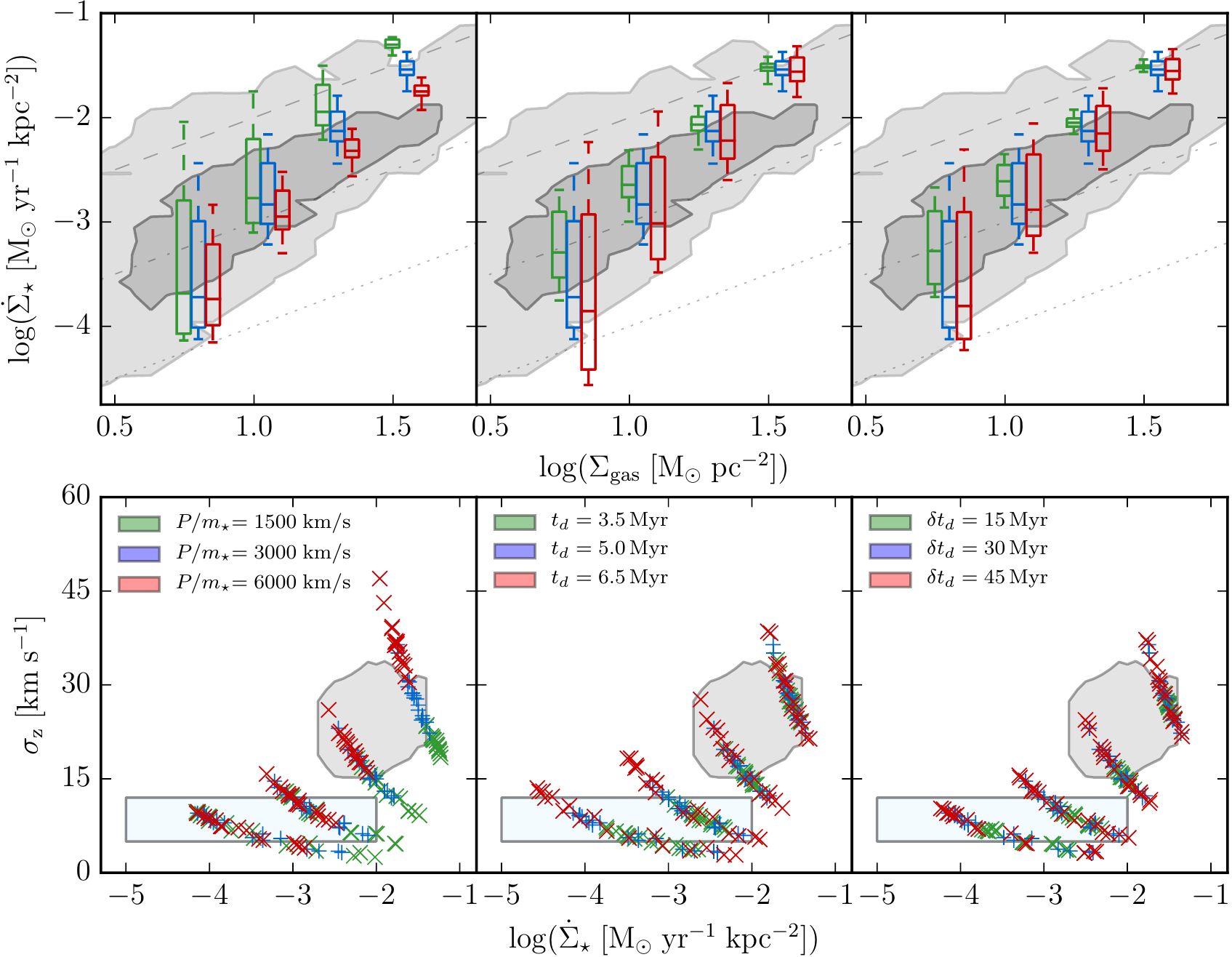}}
	\caption{Effects on the Kennicutt-Schmidt {\bf (top row)} and gas velocity dispersion--SFR {\bf (bottom row)} relations due to variations {\bf (columns)} in the overall strength ($P/m_\star$), delay time ($t_d$) and duration ($\delta t_d$) of SNe feedback in the fiducial model for $3<t\Omega<8$.  Background shaded regions (observations) and dashed lines (constant depletion times) are in the style of Figure~\ref{fig:fidKS}.  {\bf (Top row)} Box-and-whiskers for the model at a given $\Sigma_g$ are offset from the central value to show differences between model parameters; {\bf (bottom row)} colored points are sampled time-points from models at a given $\Sigma_g$, but no offsets are introduced. {\bf (Left)} Raising (lowering) the overall strength of feedback per mass of stars formed, $P/m_\star$, systematically lowers (raises) the peak/integrated star formation rates in the KS relation and raises (lowers) the gas velocity dispersion distribution at a given $\dot\Sigma_\star$.  Scatter in SFRs are also inversely affected.  {\bf (Middle)} The delay timescale before the first SNe feedback is injected, $t_d$, is a strong factor in determining the departures from SF equilibrium and their magnitudes. Longer delays produce larger departures from equilibrium.  {\bf (Right)} Varying the period over which SNe momentum is injected by a single stellar population, $\delta t_d$, affects the responsiveness of feedback to local ISM conditions.  Longer durations weaken the ability of feedback to respond quickly to the ISM conditions, resulting in more scatter in SFRs at constant $\Sigma_g$. }\label{fig:varFB_SNe}
\end{figure*}

Figure~\ref{fig:varFB_SNe} explores the effects on this model due to variations in the strength, delay time, and duration of feedback.

\begin{figure*}
	\centering
	\includegraphics[width=0.95\textwidth]{\fig{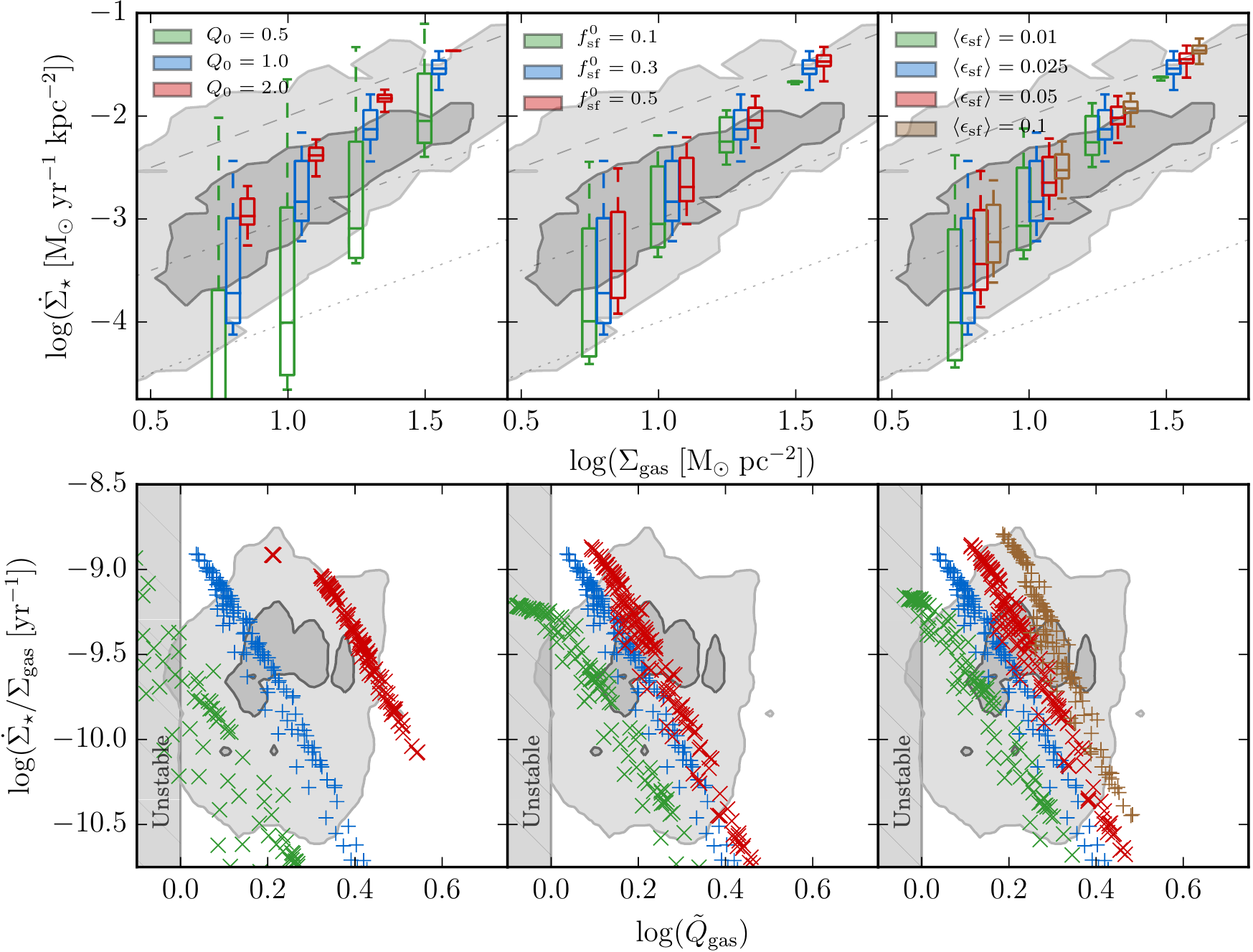}}
	\caption{Effects on the Kennicutt-Schmidt {\bf (top row)} and depletion time--stability {\bf (bottom row)} relations due to variations {\bf (columns)} in the Toomre-Q threshold ($Q_0$), maximal star-forming phase fraction ($f_{\rm sf}^0$), and average local star formation efficiency ($\left< \epsilon_{\rm sf} \right>$).  Plotted quantities and observational data regions are in the style of Figure~\ref{fig:varFB_SNe}.  {\bf (Left)} Shifting $Q_0 = 1\rightarrow 2$ moves the distributions in depletion time---stability space by $\sim 0.3$ dex, effectively renormalizing the velocity dispersions for an otherwise-constant KS relations. The scatter in SFR grows with smaller $Q_0$; as feedback injection accounts for a larger fraction of the ISM momentum budget (normalized by $Q_0$), and star formation episodes are less stable cycles than explosive events (see \S \ref{res:low}).  {\bf (Middle)} Varying the maximum fraction of gas in the star-forming phase $f_{\rm sf}^0$ is largely unimportant to the KS relation, as long as it does not ``choke" the amount of gas that would otherwise enter the star-forming phase, but shifts distributions in depletion time---stability space: lower maximum star-forming fractions require lower values of $\tilde Q_{\rm gas}$ (i.e. higher gas densities) to achieve the same SFR. {\bf (Right)} Higher local star formation efficiencies $\left< \epsilon_{\rm sf} \right>$ steepen the peak SFRs in the KS relation and shift the distributions in depletion time--stability space (higher efficiencies mean smaller quantities of unstable gas yield the same SFR), and appear to reduce scatter in KS.}\label{fig:sf_law}
\end{figure*}

\subsubsection{Feedback Strength $P/m_\star$}\label{res:vfb}

The left column of Figure~\ref{fig:varFB_SNe} shows the effects of varying the overall strength of feedback, $P/m_\star$, in our fiducial model: we plot both the Kennicutt-Schmidt relation (relating gas mass and star formation rate surface densities) and the gas velocity dispersion--SFR relation.  As demonstrated extensively in previous works exploring the feedback-regulated regime, variation in the overall strength of feedback primarily effects the equilibrium star formation rates where gas self-regulates: stronger (weaker) feedback yields lower (higher) overall star formation rates \citep{Hopkins2011, Hopkins2012a, Shetty2012a, Agertz2013, Hopkins2014, Orr2018}.  By construction, this model follows this paradigm.   %For $\Sigma_g > 20$ M$_\odot$ pc$^{-2}$, somewhat larger scatter is driven by stronger feedback, in addition to affecting the overall normalization of the star formation rate distributions.  It does affect the minimum velocity dispersions (Toomre-Qs) achieved, as lower feedback strengths take longer to arrest and reverse the run-down of turbulence and disk scale heights (hence elevated star formation overall).
Interestingly, stronger feedback (per mass of young stars) appears to result in smaller scatter in star formation rates.  As the star formation timescales, and the absolute magnitude of momentum injected by feedback, are held roughly constant between models, this can be explained as keeping the relative variance in turbulence constant across the star formation cycles.  Hence, if $\Delta \sigma \propto P/m_\star \Delta\dot\Sigma_\star$, stronger feedback produces smaller variance in turbulence for smaller variance in $\dot\Sigma_\star$.

At low star formation rates, the model is not strongly constrained to high or low feedback strengths by the spiral galaxy HI velocity dispersion dataset of the THINGS survey \citep{Ianjamasimanana2015}.  However, the higher-SFR, higher-velocity dispersion data from \citet{Zhou2017} do constrain this model in the $P/m_\star \sim 3000-6000$ km/s range.

\subsubsection{Feedback Delay Time $t_d$ and Duration $\delta t_d$}\label{res:t_d}

The middle and right columns of Figure~\ref{fig:varFB_SNe} show the effects of varying the delay timescale $t_d$ for the first SN feedback (i.e. the lifetime of the most massive star formed in a star formation event, plus the time required to propagate the SNe remnant into the ISM and drive turbulence), and the duration of SN feedback $\delta t_d$ (i.e. the difference in stellar lifetimes between the least and most massive stars to undergo a Type II SN in a star formation event).  The scatter in star formation rates is directly affected by the delay time $t_d$, with shorter delays producing less scatter in star formation rates.  Longer delay times allow for gas to over-produce stars to a greater extent before feedback is felt, hence larger departures from star formation equilibrium.  Physically reasonable values of $t_d \sim 4-6$~Myr, with a $t^{-0.46}$ weighting, are generally capable of driving $\gtrsim$dex variations in star formation rates.

In a similar vein, shorter feedback durations, $\delta t_d$, cause effectively burstier overall feedback and, as such, drive larger scatters in star formation rates.  For reasonable feedback durations of $\sim$30 Myr (roughly the difference between the lifetimes of an 8 M$_\odot$ and 40 M$_\odot$ star) the model converges on $\sim$dex scatter in star formation rates.  Longer durations smooth out feedback to the extent that it is equivalent in effect to lowering the overall strength of feedback $P/m_\star$.

\subsection{Variations in Star Formation Rate Model}\label{res:Mdot}

To bake a str\"udel, one must first cook the filling.  Analogously, in order to generate stellar feedback in a model, one must first produce stars.  The local star formation rate implemented in this model, Eq.~\ref{eq:sfr_Q}, has two principle components that we investigate.  Namely, the gas fraction in the star-forming phase $f_{\rm sf}(\tilde Q_{\rm gas};Q_0, f^0_{\rm sf}, \beta)$ (Eq.~\ref{eq:logeff}), and the average local star formation efficiency per free-fall time $\left< \epsilon_{\rm sf} \right>$.

%As detailed modeling of the dynamic chemical state of the ISM is clearly beyond the scope of this work, we simply adopt a fit to the molecular fraction by \citet{Leroy2008}, based on a previous works by \citet{Wong2002} and \citet{Blitz2006}, relating $f_{\rm H_2}$ to the ISM mid-plane pressure.  We explore the differences between this model, and another fit by \citet{Krumholz2011} in Appendix~\ref{appendix:fH2}.  By and large, the specific choice between these two models of the functional dependence of $f_{\rm H_2}$ on $\Sigma_g$ and $\Sigma_\star$, does not greatly affect the interpretations and conclusions drawn here.  The dynamical evolution of $f_{\rm H_2}$ in star forming regions of galaxies is quite interesting in its own right, having produced a large number of works in the field, and warrants future investigation in non-equilibrium models and simulations of star formation.

Varying the star formation model (i.e. the local efficiency of star formation and the Toomre-Q threshold for the onset of gravitational fragmentation/star formation) has larger systematic effects on the results of our model in depletion time--stability space compared to the effects of reasonable variations in the feedback implemented demonstrated in the previous subsection.  %In particular, the local efficiency of star formation affects the extent to which star forming gas can over-produce stars before feedback is felt, driving scatter in the SFRs observed.  Whereas, the maximal GMC fraction is less relevant so long as it is not the limiting factor in setting the SFRs.  %The steepness of the Toomre-Q threshold, $\delta Q$, affects whether or not star formation oscillates near or far from equilibrium values achieved at the Toomre-Q threshold $Q_0$ itself.

\subsubsection{Toomre-Q Threshold for Star Formation $Q_0$}\label{res:Q_0}
The left column of Figure~\ref{fig:sf_law} demonstrates the effects of the particular choice of the Toomre-Q threshold $Q_0$ on the Kennicutt-Schmidt and depletion time--Toomre-Q relations.  For physically reasonable values, the threshold sets the values of the equilibrium velocity dispersions that the models oscillate about and thus the average magnitude of turbulent momentum in ISM.  %All else being equal, the peak SFRs are not greatly affected, as peak star formation rates are dynamically limited by $t_ff \sim \sigma^{1/2} \sim Q_0^{1/2}$ and $f_{\rm sf}$ (the latter being effectively unchanged, as the behavior of $\tilde Q_{\rm gas}$ evolves dynamically with changes to $Q_0$).
Along with the overall strength of feedback, the value of the gravitational instabilities threshold is the parameter that most strongly affects the normalization of the Kennicutt-Schmidt relation in our model.

Larger values of $Q_0$ produce less scatter in the Kennicutt-Schmidt relation, as $Q_0$ sets the overall amount of turbulent momentum in the ISM ($P_{ turb,0}\sim \Sigma_g \sigma(\tilde Q_{\rm gas}=Q_0)$) where star formation occurs and thus dictates the extent to which star formation events can perturb the ISM at a given $\Sigma_g$ (see \S~\ref{res:low} for more rationale).  When $Q_0 =0.5$, the model breaks down, as feedback is able to at least double the momentum in the ISM after every star formation episode.  For values of $Q_0$ where the model holds reasonably well ($Q_0 \gtrsim 1$), doubling $Q_0 = 1 \rightarrow 2$ produces an expected $\sim 0.3$ dex shift in the Toomre-Q distribution without greatly affecting depletion times (beyond a slight tightening of the SFR distribution): gas is still able to self-regulate \citep[c.f. the predictions of][]{Krumholz2016}.

As $Q_0 \approx 1$ is a physically motivated value for the local gravitational stability threshold of the ISM \citep{Toomre1964}, and that other similar formulations of stability parameters differ only by a order-unity factor in their thresholds for gravitational fragmentation \citep{Rafikov2001,Kim2007}, we explore only a range in $Q_0$ of $0.5 - 2$.  Generally speaking, this is not a new constraint on $Q_0$, but rather shows the physical effect of varying the equilibrium level of turbulence on this non-equilibrium model (a ``robustness check" of sorts).

\subsubsection{Variations in the Maximum Star-forming Fraction $f^0_{\rm sf}$}\label{res:fGMC}

In this model, we consider that at the onset of disk scale height gravitational instabilities ($\tilde Q_{\rm gas}  = Q_0$), there is a maximum mass fraction $f_{\rm sf}^0$ of the ISM participating in star formation.  Such a constant has been adopted before in analytic models of feedback regulation in disks \citep{Faucher-Giguere2013}.  As seen in the middle column of Figure~\ref{fig:sf_law}, we see that so long as this factor $f_{\rm sf}^0$ does not `choke' the fraction of material in the star-forming phase, variations have rather small effects qualitatively.  This `choking' appears to occur at high gas surface densities where choices of small maximal fractions $\sim 0.1$ clip the maximum SFRs achieved, whereas larger values of $f_{\rm sf}$ do not appear to be the limiting factor on setting maximal SFRs (see the abrupt flattening of $f^0_{\rm sf}=0.1$ points in Figure~\ref{fig:sf_law} at short depletion times).  Larger values of $f_{\rm sf}^0$ move the distributions in depletion time--stability space to shorter depletion times and higher Toomre-Q values; this is the result of renormalizing the ``leakage" curve the model follows as $\tilde Q_{\rm gas}$ evolves (Eq.~\ref{eq:logeff}).

\subsubsection{Variations in Instantaneous Star Formation Efficiency $\left< \epsilon_{\rm sf} \right>$}\label{res:sf_eff}
%Greatly argued about in star formation literature, from GMCs on up to cosmological zoom-ins, depending on the scales considered and particular proxies for star formation observed, claimed values of the local (often interpreted as maximal) star formation efficiency fall between one or a few percent and 100\% in the conversion rate of gas to stars per local free-fall time \citep{Heiderman2010, Lada2010, Murray2011}\footnote{see \citet{Lee2016} for an investigation into the variance in $\epsilon_{\rm sf}$ finds in studies of the Milky-Way.}.  

The right column of Figure~\ref{fig:sf_law} shows how variations from $\left< \epsilon_{\rm sf} \right> = 0.01$ to $\left< \epsilon_{\rm sf} \right> = 0.1$, motivated by observational bounds \citep{Lee2016}, affect the Kennicutt-Schmidt relation, and gas depletion times and stability (Toomre-Q).  Interestingly, variations in the local efficiency over a dex change the maximal star formation rates by $\lesssim 0.5$~dex.  In the feedback regulated regime\footnote{See \citet{Semenov2018} for a recent discussion of the relative differences between feedback-regulated and dynamics-regulated star formation.}, so long as the local efficiency factor is above that required to produce \emph{enough} stars to inject the appropriate amount of feedback in the ISM to achieve equilibrium, $\left< \epsilon_{\rm sf} \right>$ should not affect the large-scale, time-averaged star formation rates.  However, lower star formation efficiencies do mean that gas must collapse to higher surface densities (i.e. reduced free-fall times) to counteract smaller local efficiencies in order to maintain the momentum balance.  More, as the gas collapses further, but does not produce more momentum in feedback overall (to first order), the distributions in depletion time--stability space shift, requiring a less stable ISM generally to support the same SFRs with lower star formation efficiencies (moving by $\sim 0.3$ dex in $\tilde Q_{\rm gas}$ for a dex change in $\left< \epsilon_{\rm sf} \right>$).

Though the effect appears less pronounced at high $\Sigma_g$, for $\Sigma_g \lesssim 10$ M$_\odot$ pc$^{-2}$, lower local star formation efficiencies produce larger scatter in star formation rates.  This is in part due to the increasing steepness of the unstable gas fraction $f_{\rm sf}(\tilde Q_{\rm gas})$, and the ability of gas to overshoot equilibrium star formation rates as the arresting effects of feedback are not felt in sufficient amounts at higher velocity dispersions (i.e., larger $\tilde Q_{\rm gas}$'s).

Given the degeneracy of the effects of variations in local star formation efficiency and the strength, delay and duration of feedback, on the Kennicutt-Schmidt relation, that relation may not be   a sensitive probe of smaller scale star formation efficiency.   Instead, observations in depletion time--stability (Toomre-Q) space have a greater ability to distinguish between low and high local star formation efficiencies in the framework of feedback regulation.  Given the definitional difficulties of a star formation efficiency in this model (i.e., that $f_{\rm sf}$ and $\left< \epsilon_{\rm sf} \right>$ could be defined together), measurements of the depletion time--stability relation \emph{in similar patches} of the ISM may be useful in quantifying ``the maximally-participating fraction" of the ISM in star formation events.  To that end, given our fiducial assumption of $f_{\rm sf} = 0.3$, our model favors low cloud-scale average star formation efficiencies $\left< \epsilon_{\rm sf} \right> \sim 0.01-0.1$, as the depletion time--stability constraints otherwise exclude $\left< \epsilon_{\rm sf} \right> \gtrsim 0.1$ for our fiducial model.

\subsection{Reproducing Resolved Galaxy Relations}\label{res:fakegal}
\begin{figure}
	\centering
	\includegraphics[width=0.42\textwidth]{\fig{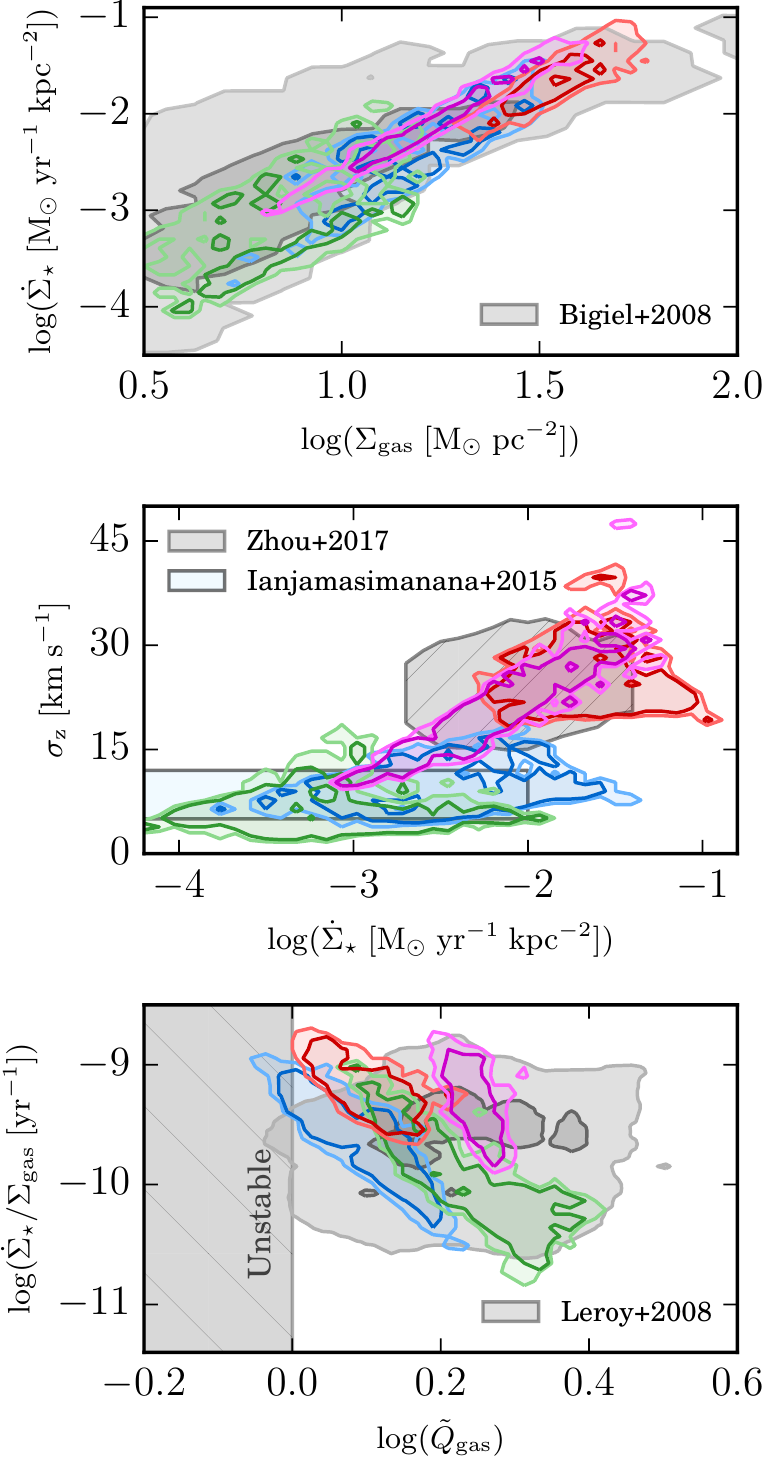}}
	\caption{Comparison of the KS, gas velocity and Toomre-Q distributions of the non-equilibrium model (brightly colored shaded regions) drawn from mock galaxies.  Plotted quantities and observational data contours are in the style of Figure~\ref{fig:fidKS}.  Mock galaxies are exponential profiles of gas and stars\rev{, whose properties are summarized in Table~\ref{table:mocks}.}  The galaxies are sampled at 750 pc resolution for radii $5 < R < 17 $ kpc, and a random time-point is chosen in the $3 < \Omega t < 8$ range for the non-equilibrium model with those local conditions.  Dark and light shaded regions indicate 50 and 90 \% inclusion regions for the model pixel distributions.  Mock distributions have significant overlap with observations in each panel\rev{, and together tile a significant portion of the observational data with modest changes in galaxy properties and star formation efficiencies.}} \label{fig:m12i}
\end{figure}

\begin{table}\caption{Properties of Mock Galaxies for Figure~\ref{fig:m12i}}\label{table:mocks}
\begin{tabular}{cccccc}
\hline
Mock & $\Sigma_{g,0}$ & $\Sigma_{\star,0}$ & $R_{g}$  & $v_c$ & $\left< \epsilon_{\rm sf} \right>$ \\
Galaxy & (M$_\odot$/pc$^{2}$) & (M$_\odot$/pc$^{2}$)  & (kpc)  & (km/s) &   \\
\hline
Blue & 100 & 1000 & 6  & 300 & 0.01  \\
Green & 50 & 500 & 6  & 300 & 0.025  \\
Red & 125 & 800 & 10  & 275 & 0.025  \\
Purple & 125 & 1000 & 6 & 290 & 0.075  \\
\hline
\multicolumn{6}{l}{Notes: $\Sigma_{g,0}$ \& $\Sigma_{\star,0}$ are central gas and stellar surface}\\
\multicolumn{6}{l}{densities for exponential disks, with scale lengths $R_g$ \& $R_\star$. }\\
\multicolumn{6}{l}{$R_\star =3$ kpc for all mock galaxies. $v_c$ is the (flat) circular}\\
\multicolumn{6}{l}{velocity, used for $\Omega = v_c/R$. $\left< \epsilon_{\rm sf} \right>$ is varied within}\\
\multicolumn{6}{l}{observational bounds $\sim 0.01-0.1$ \citep{Lee2016}.}\\
\hline
\end{tabular}
\end{table}

So far we have considered the star formation cycles of only individual patches of gas.  Given that local galaxies ($z\lesssim0.1$), unlike their high-$z$ progenitors, cannot be modeled as a single star-forming H{\scriptsize II} region, we build a snapshot of a star-forming galaxy with our model by sampling many patches of a gas disk to understand the global distribution of star formation rates and velocity dispersions.  We consider here a \rev{few} exponential disk\rev{s} of gas and stars. \rev{Table~\ref{table:mocks} summarizes the properties of these toy galaxies.}  We then discretize these disks into cartesian grids of 750 pc-sized pixels, extending 24 kpc on a side, sampling their surface densities at their centers.  For each of these points, we run our model with our fiducial parameters (see Table~\ref{table:fidmodel}), \rev{except for the cases where we have varied the small-scale star formation efficiency $\left< \epsilon_{\rm sf} \right>$,} and randomly sample one time-step to find our star formation rates, gas surface densities and velocity dispersions.  \rev{In two cases here, to highlight galaxy to galaxy variation in GMC properties, we have chosen to vary the small-scale star formation efficiency within the bounds of observations \citep{Lee2016}.  Ignored here, too, is the variance in $\Sigma_g$ at constant radius (e.g., spiral arm features) that may contribute to variance in SFE \citep{Gallagher2018}.} The results of this are seen in Figure~\ref{fig:m12i}, where we plot the resulting Kennicutt-Schmidt, gas velocity dispersion--SFR, and depletion time--Toomre-Q relations.  We compare our model mock galaxy distributions (light \& dark colored shaded regions) with resolved galaxy observations like previous figures, and find good agreement between this simple model and data.  To enable comparison, the central surface densities, scale-lengths, and orbital velocities used in our mock galaxy model were chosen to be comparable with Milky Way-mass spiral star-forming galaxies.  We do not plot pixels in our model with $R<5$~kpc, as these regions are unlikely to be modeled correctly as independent patches following cycles in star formation rate-gas velocity dispersion space (c.f., the central molecular zone of the Milky Way), given the omission of various dynamical effects like gas migration and cloud-cloud collisions \citep{Semenov2017,Semenov2018}.

The Kennicutt-Schmidt relation produced by our models in this way find good agreement with the `regulated disk' regime of \citet{Bigiel2008}.  These models produce a floor in velocity dispersions as a function of star formation rate that is somewhat lower at higher star formation rates ($\gtrsim 10^{-2}$ \mpcpc) that is somewhat below \citet{Zhou2017}.  However, given the simple structure of our mock galaxies, it is unclear if this a matter of the dynamical times or ratio of thin to thick stellar disk components being unrealistic, or a problem with the model.  Moreover, the general scatter in velocity dispersions agrees with that of the observations, using reasonably-inferred parameter values.  Lacking outflows, or some sub-grid model for local ISM heating, this model may not correctly capture the leading-edge (in SFR for a given $\sigma_z$) of the velocity dispersion relation, where the ISM can be disrupted by outflow events.

Observing the depletion time--stability relation of the mock galaxies in the bottom panel of Figure~\ref{fig:m12i}, the variations (radially) across and between the galaxies affect the normalization of the star formation-turbulence cycles of the individual patches.  This results in a widening of the relation within each galaxy, as observed on $\sim$kpc scales.  \rev{Galaxy to galaxy variations in gas and stellar properties, and variations in the star formation efficiencies of GMCs, cause the pixel distributions from the mock galaxies to tile the observational space.}  Though there is still a correlation between the quantities as observed in a single mock galaxy, the correlation is much weaker taken on the whole.  Observationally, this may present difficulties in producing a depletion time--stability relation, given that galaxy to galaxy variations in dynamical time and ratios of gaseous and stellar disk scale lengths will result in each galaxy distribution having slightly different normalizations in the depletion time--stability plane, smearing out the signal further through stacking.

\rev{Spatially-resolved observations of an individual galaxy may indeed see fairly tight correlation between depletion time and Toomre-Q, the exact slope and normalization of which will depend on the disk structure and GMC properties (here, assumed to be related to the `small-scale' star formation efficiency $\left< \epsilon_{\rm sf} \right>$).  However, this is assuming that the star formation parameters are not changing significantly across individual galaxies, e.g., small-scale star formation efficiencies having gas surface density dependencies \citep{Grudic2018}, and again that there are not significant variations in $\Sigma_g$ between independently evolving ISM patches at constant galactocentric radius.}

Non-equilibrium star formation rates, therefore, appear to produce an avenue for explaining $\sim$1-dex scatter in star formation rates in the Kennicutt-Schmidt relation, and scatter in the spatially-resolved gas velocity dispersion--SFR relation.  And although dynamical evolution of star-forming patches may obscure the relation between depletion time and stability somewhat, the variations in the disk properties across and between galaxies are more likely the reason for difficulties observing a tight correlation between Toomre-Q and SFRs \citep{Leroy2008}.

%------------------------------------------------------------------------------------
\section{Discussion}\label{sum}

\subsection{The ``Instantaneous" Feedback Timescales Limit}\label{sec:instSFReff}

Much of this work focuses on the case where the feedback delay timescales $t_d$ and $t_d + \delta t_d$ are within an order of magnitude of the local dynamical time of the galaxy $1/\Omega$ (or for strongly self-gravitating disks, $1/\sqrt{4 \pi G \rho_0}$).  In the case where $t_d$ and $\delta t_d \ll 1/\Omega$, however, star formation and feedback can be treated as occurring ``instantaneously" after a delay time $t_d$, compressing all SNe and prompt feedback into a spike at $t_d$.  We too can consider the case when the star formation threshold is very sharp, i.e. $\beta \rightarrow \infty$ such that Eq.~\ref{eq:logeff} becomes
\begin{equation}\label{eq:stepeff}
f_{\rm sf}(\tilde Q_{\rm gas}) =  \theta(Q_0 - \tilde Q_{\rm gas}) f^0_{\rm sf} \; ,
\end{equation}
where $\theta(Q_0-\tilde Q_{\rm gas})$ is the Heaviside step function at the star formation threshold of $\tilde Q_{\rm gas}=Q_0$.  In this setting, the turbulent velocity dispersion $\sigma$ is not allowed to fall much below the threshold value at $Q_0$, since feedback acts effectively instantaneously once star formation begins to occur.

Thus, the amount of star formation that occurs in a star formation episode is just the amount that can form in one feedback timescale.  So, we form an amount of stars per event
\begin{equation}\label{eq:deltaMstar}
\Delta \Sigma_\star = \left< \epsilon_{\rm sf} \right> f^0_{\rm sf}\Sigma_g t_d /t_{\rm eddy} \; .
\end{equation}
Interestingly, the amount of stars formed has no (direct) relation to the absolute strength of feedback, so long as the amount of momentum eventually injected back into the ISM from this mass of stars is enough to at least momentarily halt additional star formation.  The time between star formation events is dependent on the fact that each event will pump up the turbulent velocity dispersion by $\Delta \sigma = (P/m_\star)\Delta \Sigma_\star / \Sigma_g$.  This extra momentum, above that required strictly to maintain stability, takes a time $t_{cycle}$ to decay back down to the star formation threshold $\sigma(\tilde Q_{\rm gas} = Q_0)$ of
\begin{equation}
t_{cycle} = \ln({1 + \Delta \sigma / \sigma(\tilde Q_{\rm gas}=Q_0)})/\Omega \; . 
\end{equation}
It is worth noting, that for the outskirts of galaxies, where the quantity $t_d \Omega$ is likely to be small as we assumed ($1/\Omega$ being the dominant component of the local dynamical time, thanks to exponentially falling disk surface densities), galaxy disks are seen to have relatively constant H{\scriptsize I} disk velocity dispersions \citep{Tamburro2009}, and so we expect the ratio of $\Delta\sigma/\sigma(\tilde Q_{\rm gas}=Q_0)$ to be small.  Thus, we can approximate $t_{cycle}$ as $t_{cycle} \approx \Delta \sigma / \sigma(\tilde Q_{\rm gas}=Q_0) \Omega $. 

And so the average star formation rate over a star formation cycle\footnote{This is identical to averaging it over a dynamical time, as then we have a star formation rate of $\Delta \Sigma_\star \Omega / \Omega t_{cycle} = \Delta \Sigma_\star / t_{cycle}$.} is $\bar{\dot \Sigma}_\star = \Delta \Sigma_\star / t_{cycle}$.  Explicitly,
\begin{equation}
\bar{\dot \Sigma}_\star \approx \frac{ \Sigma_g \Omega \sigma(\tilde Q_{\rm gas} = Q_0)}{P/m_\star } \; .
\end{equation}
The average efficiency of star formation per dynamical time is then
\begin{equation}
\bar \epsilon_{\rm sf} = \frac{\bar{\dot \Sigma}_\star}{\Sigma_g\Omega} \approx \frac{ \sigma(\tilde Q_{\rm gas}=Q_0)}{P/m_\star} \; .
\end{equation}
Neither the average star formation rate nor the average star formation efficiency have an explicit dependence on the ``small scale" (GMC-scale) star formation efficiency (here, $\left< \epsilon_{\rm sf} \right>$) or eddy-crossing/free-fall time $t_{\rm eddy}$, or feedback delay timescale $t_d$ (provided $t_d \Omega \ll 1$), so long as the amount of stars formed in a star formation episode injects enough momentum to regulate the ISM but not enough to fully disrupt it (i.e. drive $\tilde Q_{\rm gas}$ to $\gg 1$).  Unsurprisingly, this is identical to the result of \S~\ref{ss:eq_sfr}, though we are considering a case of extreme dis-equilibrium. This is complementary to the picture of feedback regulation in \citet{Semenov2018}, where low star formation efficiencies produce high duty cycles of star formation- after all, less stars formed means $\Delta\sigma/\sigma(\tilde Q_{\rm gas}=Q_0)$ will be smaller.  Plugging in `typical' values for $\sigma(\tilde Q_{\rm gas}=Q_0) \approx 15-45$ km/s and $P/m_\star \approx 3000$ km/s yields a global, averaged star formation efficiency of $\bar \epsilon_{\rm sf} \approx 0.005-0.015$.  These are not altogether unreasonable values for the star formation efficiency in the outskirts of galaxies \citep{Bigiel2010}, and in agreement with the median values of star formation efficiencies of our fiducial model.  This provides a reasonable mechanism, reliant on averaging non-equilibrium star formation episodes, for regulating \emph{local} star formation (of any efficiency) to global inefficiency on galactic dynamical timescales.

\subsection{Low Gas Surface Density Regime/Limit}\label{res:low}

Seen clearly across the Kennicutt-Schmidt panels of Figures~\ref{fig:varFB_SNe} and \ref{fig:sf_law}, the delayed feedback model drives large $\sim 2$~dex scatter in SFRs for gas surface densities $\lesssim 10$~M$_\odot$ pc$^{-2}$.  As the gas surface density falls below 10~M$_\odot$ pc$^{-2}$, two processes dovetail to make our feedback regulated turbulent disk model break down.

Below $\sim 10$~M$_\odot$ pc$^{-2}$, the gas disk transitions from a supersonic(turbulently-supported) molecular disk, to a transonic atomic disk (with non-negligible thermal support), as the sound speed of 6000~K gas is almost but not quite sufficient with $c_s \sim 6$~km/s to maintain $\tilde Q_{\rm gas} \sim 1$ (i.e., providing nearly half of the required support).  In these circumstances, stirring due to supernovae no longer dominates as the sole process stabilizing the ISM on kpc-scales, and the maintenance of thermal support in a two-phase medium becomes necessary to include.  The thermal support component, and its connection to stellar feedback, is not included in the model, as it would require modeling the molecular gas fraction $f_{\rm H_2}$ and gas cooling, which is beyond the scope of this work.  Further, given the increasingly two-phase nature of the ISM at low $\Sigma_g$, the treatment of the star forming fraction $f_{\rm sf}(\tilde Q_{\rm gas})$ as a simple power law may break down, contributing to a change in kpc-scale star formation efficiencies \citep{Schaye2004, Krumholz2009, Krumholz2018}.  Additional considerations at low gas surface densities include the ability of gas self-gravity (not included) to drive sufficient turbulence in the outer H{\scriptsize I} disks \citep{Agertz2009}.

On the other hand, for the ``lightest" cold, turbulently-supported disks with surface densities $\sim 10$~M$_\odot$ pc$^{-2}$, SNe feedback from star formation events can inject significant fractions of the turbulent momentum in the disk.  Take a star formation event at a gas surface density of $10$~M$_\odot$ pc$^{-2}$, where our fiducial model reaches peaks star formation rates of $\dot\Sigma_\star \sim  10^{-2.5}$~M$_\odot$ kpc$^{-2}$ yr$^{-1}$ for $\sim10^7$ yr (c.f., plausible GMC lifetimes) producing $\sim 10^{4.5}$~M$_\odot$ kpc$^{-2}$ of stars. These young stars then result in a SNe density of $\sim 10^{2.5}$~kpc$^{-2}$ in the proceeding $\sim 40$~Myr \citep[given a rate of a single SNe per 100 M$_\odot$ of stars formed;][]{Ostriker2010}.  At a momentum per Type-II SNe of $\sim 3 \times 10^5$~M$_\odot$ km/s \citep{Martizzi2015}, this is a turbulent momentum injection of $\sim 10^8$~M$_\odot$~km/s~kpc$^{-2}$.  For a  $\sim 10$~M$_\odot$ pc$^{-2}$ gas disk, with $\tilde Q_{\rm gas}\sim 1$ ($\sigma \sim 10$~km/s), the total turbulent gas momentum is $\sim \Sigma_g \sigma(\tilde Q_{\rm gas}\sim1) \sim 10^8$~M$_\odot$ km/s kpc$^{-2}$.  As the momentum injected is a non-negligible (tens of percent approaching unity, with uncertainty regarding the feedback budget per SNe \citealt{Fielding2018}, \citealt{Gentry2019}) fraction of the momentum contained in the turbulence field of the whole disk patch, feedback is increasingly disruptive to the disk structure. This is more or less the difference between SNe clusters blowing holes in the ISM (dominating), versus churning or stirring it (perturbations).

And so, given that our model does not capture the feedback, star formation and gas physics of the transition from a predominantly-atomic ISM with non-negligible thermal support to a turbulently-supported, molecularly-dominated one, this model exhibits increasingly disruptive star formation events at low gas surface densities.  It is not clear, on the basis of this model alone, the extent to which growing scatter ($\gtrsim 2$ dex) in star formation rates due to the time-lag of feedback injection are to be expected for low ($\lesssim 10$ M$_\odot$ pc$^{-2}$) gas surface density regions.  Broadly, this is exemplary of the difficulties in modeling the variety of star formation environments within galaxies with a single, simple model.

\section{Conclusions}\label{conclusions}
In this paper, we developed a simple, non-equilibrium model of star formation in the context of sub-kpc patches of disk galaxies (c.f. local disk scale heights) and explored its ability to explain the scalings and scatter in galaxy star formation relations.  Our principal conclusions are as follows:
\begin{itemize}

\item The local strength of feedback $P/m_\star$, in addition to setting the normalization of the KS relation, itself may contribute to setting the scatter in observed SFRs.   If the variance in turbulent momentum is roughly constant through star formation events, then the variance in SFRs is inversely proportional to $P/m_\star$  through $\Delta\sigma \propto P/m_\star \Delta\dot\Sigma_\star$.

\item Longer delay times between star formation and the injection of feedback $t_d$ and overall injection intervals $\delta t_d$ are able to drive larger departures from star formation equilibrium.  This occurs because the ISM is able to ``overshoot" and over-produce stars to a greater extent, and the subsequent feedback events drive larger velocity dispersions (Toomre-Qs).  Delay times on the order of 4-6 Myr produce $\sim$dex scatter in SFRs.

\item The relative steepness of the gravitational instabilities threshold and the timescale of feedback injection may together explain the large range of SFRs seen at low $\Sigma_g$ with little variance in velocity dispersions in outer HI disk velocity dispersion profiles \citep[e.g., spiral galaxy HI disks in the THINGS survey,][]{Ianjamasimanana2012, Ianjamasimanana2015}.

\item This model predicts a correlated depletion time--Toomre-Q relation for individual galaxies (c.f., bottom panel of Figure~\ref{fig:m12i}).  However, within individual galaxies a degree of scatter is introduced as the normalization and slope of the locally tightly evolving relation varies across disks with the changing disk properties.  Further smearing of this relation is introduced in galaxy surveys by stacking different galaxies with altogether different disk and GMC properties (with their attendant differing slopes and normalizations of the depletion time--stability relation).

\end{itemize}

The proposed non-equilibrium star formation model can explain the observed $\sim1$~dex scatter in resolved star formation scaling relations.  More so than the effects of metallicity or variations in gas fraction, non-equilibrium states of star formation can explain large variations in average star formation rates (e.g. H$\alpha$-inferred SFRs).  This arises due to the fact that the interplay of bursty feedback, injected over some finite timescale, and the roughly smooth dissipation of turbulence (on $\sim$kpc scales) struggles to find a stable balance on timescales of tens of Myrs.

Careful spatially-resolved observations of \emph{individual} star forming galaxies may be able to identify a depletion time--Toomre-Q relation, provided that the effects of variations in gas fraction at constant radius and changes in star formation efficiency within GMC across the disks can be accounted for.  Indeed, the slope and normalization of this relation may even inform on the small-scale star formation efficiency within those specific galaxies.

Future work using resolved galaxy surveys, like the MaNGA and SAMI surveys, at the sub-kpc scale may help to elucidate the extent to which the scatter in resolved star formation rates correlates with dynamical conditions at the disk scale.  The ability to marshal statistically significant samples of star-forming regions with similar physical conditions may make it possible to disentangle potentially confounding local quantities such as metallicity or gas fraction.

%------------------------------------------------------------------------------------
\section*{Acknowledgments}
MEO is grateful for the encouragement of his late father, SRO, in studying astrophysics.  We are grateful to the anonymous referee for providing us with constructive comments and suggestions that greatly improved the quality of this work.  MEO is supported by the National Science Foundation Graduate Research Fellowship under Grant No. 1144469.  The Flatiron Institute is supported by the Simons Foundation.  Support for PFH was provided by an Alfred P. Sloan Research Fellowship, NASA ATP Grant NNX14AH35G, and NSF Collaborative Research Grant \#1411920 and CAREER grant \#1455342.  This research has made use of NASA's Astrophysics Data System.

%   Numerical calculations were run on the Caltech compute cluster ``Zwicky'' (NSF MRI award \#PHY-0960291) and allocations TG-AST120025, and TG-AST130039 granted by the Extreme Science and Engineering Discovery Environment (XSEDE) supported by the NSF.  CAFG was supported by NSF through grants AST-1412836 and AST-1517491, and by NASA through grant NNX15AB22G. DK acknowledges support from NSF grant AST-1412153 and the Cottrell Scholar Award from the Research Corporation for Science Advancement.  DMS is supported by the National Science Foundation Graduate Research Fellowship under Grant No. 2015192719.

\bibliography{/Users/mattorr/Desktop/Hopkins/Docs/library,alt_bibs}
\bibliographystyle{mnras}

\appendix

\section{Parameters of Supernova Feedback}\label{append:SNe}
The lifetimes of massive (8-40 M$_\odot$) stars that are the progenitors of Type II SNe events are fairly well constrained for our purposes.  Furthermore, the slope of the massive end of the stellar initial mass function (IMF) is also well known \citep[see][and references therein]{Krumholz2014,Offner2014}.  Together, these constraints put a strong prior on the parameter space to be explored by this model, in terms of the delay time to the first effects of SNe feedback being felt, how long feedback events last, and the relative distribution of feedback injection in time after a star formation event.

From stellar evolution theory, the main sequence lifetimes of the most massive stars in the local universe range from approximately 4.5 to 38 Myr for 40 to 8 M$_\odot$ stars \citep{Raiteri1996}.  We take the lifetime of a 40 M$_\odot$ star as a bound for the minimum delay time to the first SNe feedback effects in our model $t_d$.  Admittedly, longer delay times by perhaps a factor of two are not unreasonable given the (un)likelihood of forming the most massive star first in a local star formation episode, in addition to the various effects rotation and binarity.  On the other hand, there is a broader absolute range in the time for the last Type II SNe to go off of 30-49 Myr (approximately factor of two uncertainty), given the uncertainty in the lower mass limit for Type II SNe progenitors of $8 \pm 1$ M$_\odot$ \citep{Smartt2009}.

To constrain the distribution in time of Type II SNe events from a star formation episode (between the most- and least-massive progenitor's endpoints), i.e., $dN_{\rm SN}/dt$, we combine the IMF slope $dN/dM_\star$ and the mass dependence of main sequence lifetimes (specifically $dt/dM_\star$).  Taking the lifetimes of massive stars to be proportional to their mass-to-light ratios $t(M_\star) \propto M_\star/L_\star$ and with $L_\star \propto M_\star^{3.5}$, we have $t(M_\star) \propto M_\star^{-2.5}$ (or $M_\star \propto t^{-2/5}$) and thus $dM_\star/dt \propto t^{-7/5}$ \citep{Boehm-Vitense1992}.  From the slope of the high-mass end of the IMF, we take the canonical Salpeter IMF slope of -2.35, i.e. $dN/dM_\star \propto M_\star^{-2.35}$, and in terms of their stellar lifetimes $dN/dM_\star$ is then $\propto t^{4.7/5}$.  Combining these arguments, we yield a power-law distribution of,
\be
\frac{dN_{\rm SN}}{dt} = \frac{dN}{dM_\star} \frac{dM_\star}{dt} \propto t^{-0.46}  \; ,
\ee
which is fairly weak (though not flat) in time, as the shorter lifetimes of the most massive stars nearly balance out with their relative rarity.

For the purposes of this study, we thus adopt an initial delay time of $t_d=$~\tdo~Myr, a feedback episode period of $\delta t_d=$~\dtdo~Myr, and a time-weighting of $dN_{\rm SN}/dt \propto t^{-0.46}$.

\section{What about SFR averaging timescales?}\label{res:times}
\begin{figure}
	\centering
	\includegraphics[width=0.42\textwidth]{\fig{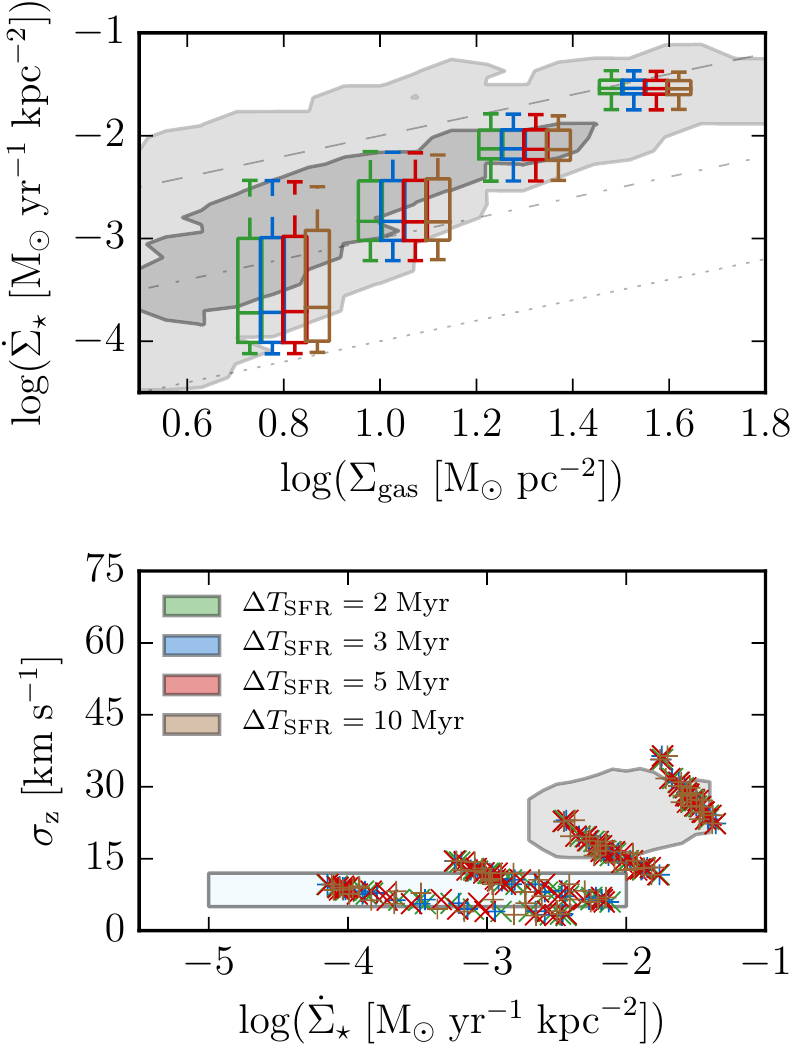}}
	\caption{Effects of variation in the star formation averaging period on the model KS and gas velocity dispersions for fiducial model parameters.  Observational (KS) data and plotted quantities are in the style of Figure~\ref{fig:varFB_SNe}.  For reasonable choices of averaging period between 2-10 Myr (c.f., the H$\alpha$ tracer timescale and timescales thereabouts), little to no effect is seen on the average star formation rate distributions. %Though longer timescales do smooth out star formation rate peaks at lower gas surface densities somewhat, as the ``on" fraction of the star formation duty cycle is the shortest there.
	} \label{fig:AvgWin}
\end{figure}

Observationally, the ``instantaneous" star formation rate of a region is ill-defined.  YSO counts are perhaps the closest proxy to an true instantaneous star formation rate, but even they have a spread in their lifetimes (hence the averaging timescale of SFRs inferred) of as little as 0.5 Myr for 0/I YSOs to being a Myr or more removed from the star formation event itself in the case of Class II YSOs \citep{Evans2014, Heyer2016}.  As such, any model of non-equilibrium star formation must be convolved with an averaging timescale for the observable tracer.  In the case of H$\alpha$ or IR flux, we are averaging over a $\sim 2-4$ Myr timescale, for tracers like the FUV flux, that timescale is significantly longer ($\sim 30$ Myr).  Hence, variability in star formation rates on timescales shorter than the averaging timescale of the particular tracer investigated will be smoothed out.  We investigate the effects of particular choices of averaging period $\Delta T_{\rm SFR}$ in Figure~\ref{fig:AvgWin}, wherein we convolve the instantaneous star formation rates produced by our model (Eq.~\ref{eq:sfr_Q}) with a 2-10~Myr wide time-averaging window $\Delta T_{\rm SFR}$. Specifically choosing this timescale to be a proxy for the H$\alpha$ and IR flux-inferred star formation rates, to show how the variations in SFR over the cycle are smoothed out.  Increasing the averaging window blunts the star formation rate maxima achieved, as the peak in the star formation cycle is smoothed to some degree.  The particular choice of averaging window does not alter the predictions of the model with respect to $\Sigma_{\rm gas}$ or $\sigma_z$.  The averaging effects on $\dot \Sigma_\star$ are relatively small as $\Delta T_{\rm SFR} \Omega \sim 0.1$ in our fiducial model, and so the averaging window constitutes only a fraction of a star formation cycle.  Throughout the main body of the text, we adopt a canonical 3~Myr averaging window for our star formation tracer for simplicity.

\label{lastpage}
\end{document}